\documentclass[12pt,preprint]{aastex}

\def\aj{{AJ}}
\def\araa{{ARA\&A}}
\def\apj{{ApJ}}
\def\apjl{{ApJL}}
\def\apjs{{ApJS}}

\def\aap{{A\&A}}

\def\mnras{{MNRAS}}

\def\qjras{{QJRAS}}

\slugcomment{Version \today;\\ 
Accepted and scheduled for ApJ 
2008, v677 n1 issue}

\shorttitle{
Continuum Imaging
of Elliptical Galaxies at 350$\mu$m} 
\shortauthors{Leeuw et al.}

\begin{document}


\title{Spatially Resolved Imaging at 350$\mu$m of Cold Dust in Nearby Elliptical Galaxies}

\author{Lerothodi L. Leeuw\altaffilmark{1,2}, Jacqueline Davidson\altaffilmark{3}, C. Darren Dowell\altaffilmark{4}, and Henry E. Matthews\altaffilmark{5}}

\altaffiltext{1}{Department of Physics \& Electronics, Rhodes University, PO Box 94, Grahamstown 6140, South Africa; lerothodi@alum.mit.edu.}
\altaffiltext{2}{Space Science and Astrophysics Branch, NASA Ames Research Center, MS 245-6, Moffett Field, CA 94035; lleeuw@arc.nasa.gov.}
\altaffiltext{3}{USRA-SOFIA, NASA Ames Research Center, MS 211-3, Moffett Field, CA 94035; jdavidson@sofia.usra.edu.}
\altaffiltext{4}{Jet Propulsion Laboratory, Mail Stop 169-506, 4800 Oak Grove Drive, Pasadena, CA  91109; charles.d.dowell@jpl.nasa.gov.} 
\altaffiltext{5}{Herzberg Institute of Astrophysics, National Research Council of Canada, P. O. Box 248, Penticton, BC, V2A 6J9, Canada; Henry.Matthews@nrc-cnrc.gc.ca.}



\begin{abstract}
Continuum observations at 350\,$\mu$m
are presented of 
seven 
nearby elliptical galaxies, for which
CO-gas disks have recently
been resolved 
with interferometry mapping.  
These SHARC\,II mapping results
provide the first {\em clearly resolved}
far-infrared(FIR) to submillimeter(submm)
continuum emission from cold dust (with temperatures 32\,K $> T >$
22\,K) of any elliptical galaxy at a distance $>40$\,Mpc.       
The measured FIR excess shows that the 
most likely and dominant heating source 
of this dust 
is not dilute stellar radiation or cooling flows, but rather
star-formation, that 
could have been 
triggered by an accretion or
merger event and fueled by dust-rich
material 
that has settled in a 
dense region 
co-spatial with the
central CO-gas disks.
The 
dust is detected even in two 
cluster ellipticals
that are deficient in H\,I, showing 
that, unlike the H\,I, cold dust and CO in ellipticals can survive 
in the presence of hot X-ray gas, even in galaxy clusters.
No dust cooler than 20\,K, either distributed outside the CO disks, 
or co-spatial with and heated by the entire dilute stellar optical galaxy
(or very extended H\,I), 
is currently evident.


\end{abstract}


\keywords{galaxies: elliptical and lenticular, cD --  galaxies: ISM
  --  galaxies: photometry -- infrared: galaxies --  radio continuum:
  galaxies -- submillimeter}


\section{Introduction}
\label{sec:intro_es}

Cold gas and dust in nearby elliptical galaxies were detected only
fairly recently 
\citep[][]{sad85, kna89, lee91, wik95}.
Compared to that in spiral galaxies, 
the cold Interstellar Medium (ISM) 
in ellipticals is  
present in relatively small amounts and is seen in only 50\% to 80\% of
nearby ellipticals.  The source and content of the cold ISM in these
galaxies is still uncertain, with optical and
far-infrared (FIR) dust-mass estimates differing by $\sim$ 10 to 100
\citep[e.g.,][]{gou95_b}.

There is some evidence for an internal origin of the dust 
arising in circumstellar envelopes of red giant stars. 
This internal origin is supported by the detection of
$\sim 10\,\mu$m emission  
in excess of the stellar emission, 
with the same de Vaucouleurs
profile as the optical stellar  
images and near- to mid-IR photospheric emission 
\citep[e.g.,][]{ath02}.
 Some have 
suggested that 
this dust would come in contact with the hot ISM ($\sim 10^7$\,K) and
be eroded in thermal collisions  
with ions and destroyed in only $\sim 10^7$ to $10^8$\,
years 
\citep[e.g.,][]{sch98_w}.
 However, 
\citet[][]{mat03}
have argued that the dust may induce efficient cooling of the hot gas
on a timescale shorter than  
the destruction lifetimes of the dust in the hot gas, allowing
some dust to survive and settle primarily in the centers of the galaxies
while the hot gas cools.  They 
propose that if
this dust-induced cooling is the dominant evolutionary path of the dust
from stars, the thermal energy released by the hot ISM should be 
re-radiated by dust in the FIR and could be responsible
for the central dusty disks and clouds seen in ellipticals. 

It has also been suggested that mergers or accretion
events that include a gas-rich galaxy may be the dominant source of dust
in ellipticals.  The total far-IR luminosities 
\citep[e.g.,][]{tem04}
and cold gas content 
\citep[][]{kna85, lee91}
 of ellipticals are observed to be
uncorrelated with their optical  
luminosities.  This lack of correlation, together with HST detections
of irregular dust lanes and
filaments with random orientations with respect to the optical major
axes 
\citep[e.g.,][]{tra01}
is usually interpreted as evidence that the dust has been externally
acquired. The external origin  
is likely
to be accretion 
via the interaction of an elliptical galaxy
with a gas-rich spiral, or in  
mergers of gas-rich spirals that evolve through an FIR-bright
(starburst) phase as they relax to  
form ellipticals 
\citep[e.g.,][]{too72, sch98_w}.  
In fact, if elliptical galaxies are formed through spiral mergers,
then starbursts are necessary to produce the high phase-space
densities seen in the centers of elliptical galaxies \citep[e.g.,][and
references therein]{rot04}. 
Numerical simulations by 
\citet[][]{bar02}
 have shown that 20\% to 60\% of the gas from two spirals
merging can form relaxed disks  
of sizes up to 20\,kpc, with some of the gas maybe losing angular momentum,
falling to the nucleus of  
the galaxy, and perhaps leading to star-formation.

Observations of dust emission from elliptical galaxies have 
been made using $IRAS$ or $ISO$ primarily   
\citep[e.g.,][]{kna89, tem04}.
  These measurements
favored the warm dust  
components of the ISM and produced unresolved images of the dust
emission.  The $IRAS$ and $ISO$  
derived SEDs of these galaxies clearly show that there are
multi-temperature components to  
the dust. Indeed, the presence of an additional, low-level, very cold
($<20$\,K) dust component in ellipticals has been suggested by limited
detections of unresolved mm and submm emission in some nearby galaxies
\citep[][]{kna91, wik95b, lee04, tem04},
 as well as clearly extended FIR to submm detections from the
dust lane in the nearest giant elliptical galaxy NGC\,5128 that hosts
the active galactic nucleus (AGN) Centaurus\,A 
\citep[e.g.,][]{lee02, qui06}.
Observations of CO emission from a number of ellipticals
show that many of them  
contain significant amounts of
molecular gas 
\citep[][]{lee91, wik95},
but many of these observations do not spatially resolve the CO emission.  

Recently, 
interferometric measurements 
by \citet[][]{you02, you05}
 and 
\citet[][]{wik97}
 have spatially  
resolved the CO (1-0) emission in seven elliptical galaxies with spatial
resolutions of 7 and 3 arcsec,  
respectively. These data imply that the molecular gas is 
contained mostly
in rotating disks lying within  
the central 5\,kpc
 of the galaxies. 
This prompted the study by
our group to spatially resolve  
the dust emission in these seven elliptical galaxies. 
Our goals were (a) to determine the physical properties of  
the dust coincident with the molecular gas; (b) to see if there are
dust associations with other  
spatial distributions (e.g., stellar distribution); and (c) to
determine the most likely heating source (or sources) for the dust.
We wanted also to address the question of whether these observed gas
disks are the result of external merger/accretion events, as suggested
by the optical morphological signatures of the galaxies, or 
internal mass-losing stars that constitute their elliptical optical
components.     

The seven galaxies in our sample (NGC\,83, NGC\,759, NGC\,807, UGC\,1503, NGC\,3656,
NGC\,4476, and NGC\,5666) 
were all classified by 
\citet[][]{wik95}
 as ellipticals, based on 
the determination that their luminosity profiles follow the $R^{-1/4}$ de Vaucouleurs law. 
However, 
morphologically
the sample 
includes galaxies 
from different extragalactic environments and represents a spread of 
possible 
merger/accretion tracers or ages,
from galaxies that have been classified as on-going or early-age 
major mergers 
\citep[e.g., NGC\,3656,][]{bal01}
 to very-late accretion or quiescent systems 
\citep[e.g., NGC\,807,][]{mur00}.
  Recent CCD imaging of some of
the galaxies, 
e.g., NGC\,5666 by
\citet[][]{don03},
have disputed the
classification as ellipticals in favor of SO galaxies, while new
optical observations of others, 
e.g., NGC\,83 by
\citet[][]{you05},
have
confirmed the elliptical classification.  


The sample's updated environments, general properties,  
and molecular gas content are summarized in Table 1.
The observational merger or accretion signatures of the sample
 are worthy of note as
they fortuitously 
allow an evolutionary analysis in the context 
of
merger/accretion 
stages 
and 
cold dust and gas properties. These merger/accretion signatures are
described further with
other observational characteristics of the  
galaxies in Section~\ref{sec:submm_morph} below.  

\section{Mapping Observations 
using SHARC\,II at 350$\mu$m}\label{sec:sharc_obs}  

We observed NGC\,83, NGC\,807, UGC\,1503, and NGC\,759 
with SHARC\,II
on 
October 11--14, 2005, 
and  NGC\,3656,
NGC\,4476, and NGC\,5666 on 
April 23 and 29, 2006.  
SHARC\,II has a $12 \times 32$
bolometer array with an instantaneous field of view of  
$148'' \times 56''$.
We achieved background subtraction and mapping coverage
for the galaxies by having the telescope execute a 
Lissajous scanning pattern with amplitude $60''$ in azimuth and
$40''$ in elevation.
During the observations, we 
maintained a log of
air mass and the zenith water vapor; the telescope dish-segments were
kept in optical alignment using the CSO Dish Surface
Optimization System (DSOS); and pointing calibrations were executed every
hour, allowing us to achieve a FWHM beamsize of $8.5''$
and pointing accuracy of about $2''$ rms.  
The zenith atmospheric opacity at 225\,GHz, during the observations
and as measured by the CSO facility radiometer, varied from 0.04 to 0.06.

For flux calibration, we observed
either Callisto, Uranus, CRL\,618, CRL\,2688, Vesta, Arp\,220, CIT6, Neptune,
or o\,Cet hourly.
The assumed fluxes for the exosolar calibrators are 
10.2\,Jy/beam for Arp\,220,   
19.4\,Jy/beam for CRL\,618,   
2.33\,Jy/beam for o\,Cet, 
41.6\,Jy/beam for CRL\,2688, and 
2.42\,Jy/beam for CIT6, and, for the solar-system calibrators, 
in October 
232\,Jy/beam for Uranus,
90.0\,Jy/beam for Neptune,  
5.77\,Jy/beam for  Vesta,
and 
in April 
95.8\,Jy/beam for Callisto.
The absolute
calibration uncertainty is estimated to be 20\%.  

\begin{figure}[!th]
\begin{center} 
\includegraphics[angle=0,height=2.518in]{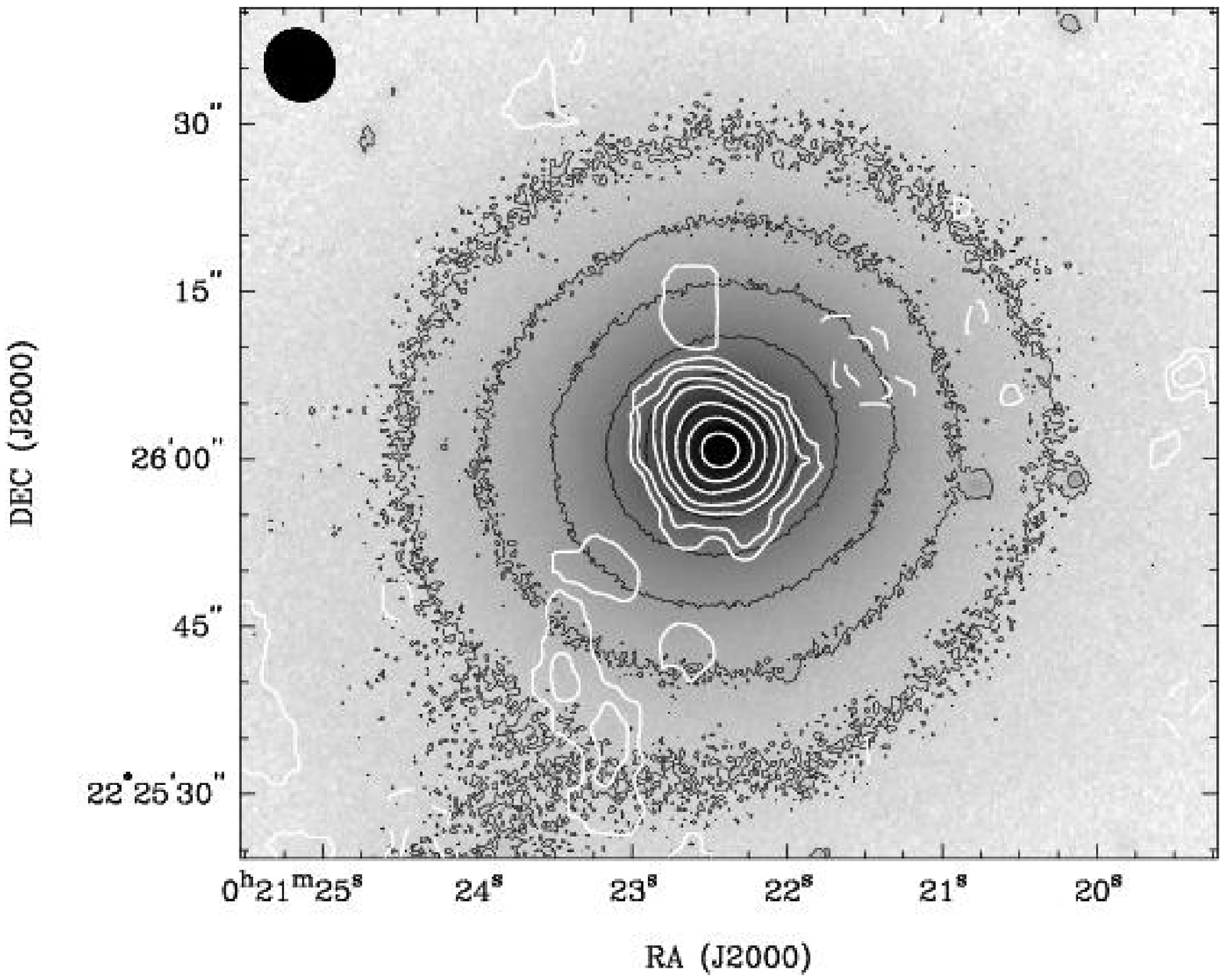}
\includegraphics[angle=0,height=2.518in]{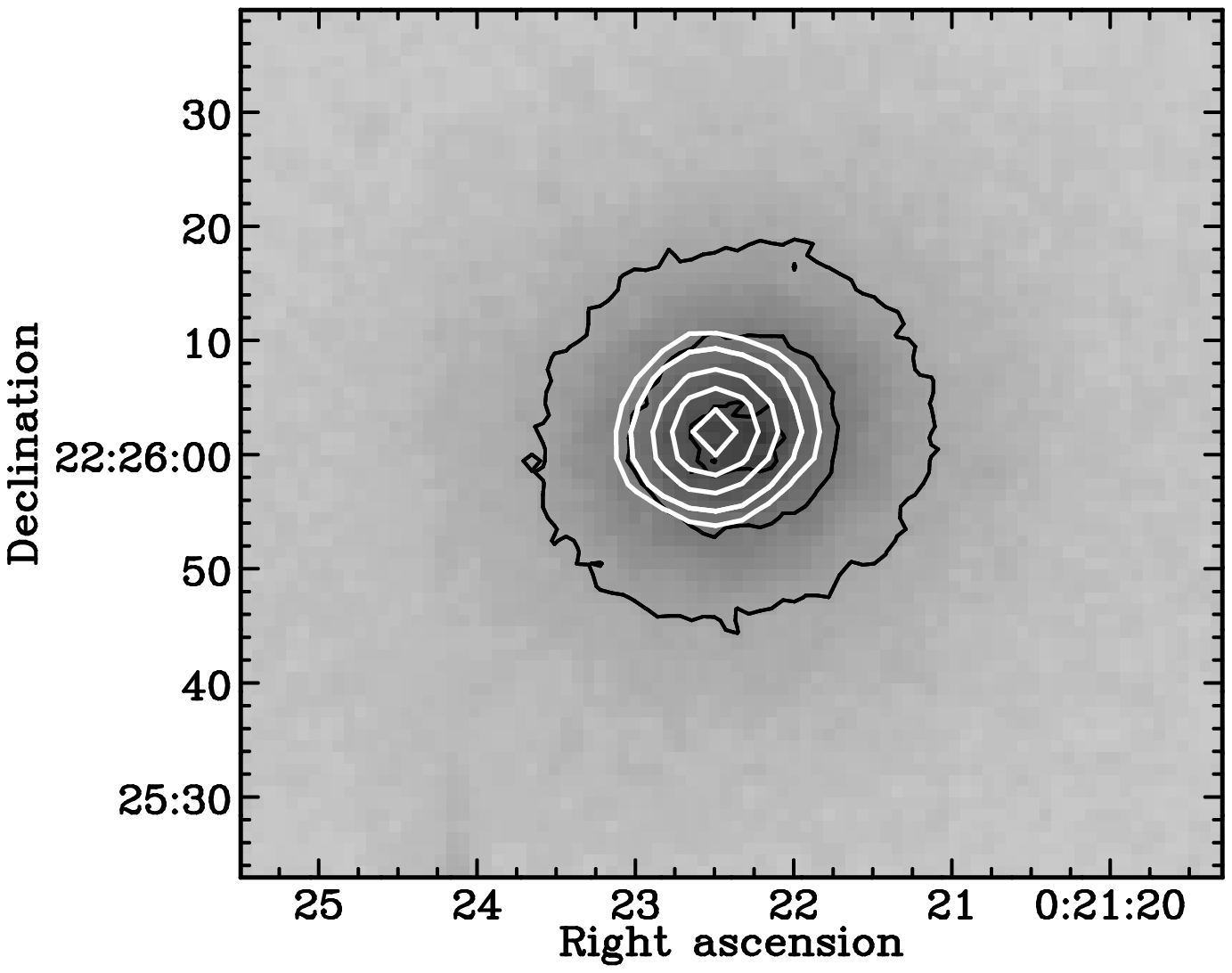}
\caption{
{\it Left:} 
White contours of the total integrated CO(1-0) intensity map 
of NGC\,83, with a measured 10\%-contour-level diameter of $10''$,
are overlayed on the grey scale and black contours 
from the 
red part of the optical Digitized Sky Survey 2 (DSS2), as presented  by 
\citet{you05}. 
The white contours are in units 
of -20\%, -10\%, 10\%, 20\%,
30\%, 50\%, 70\%, and 90\% of 
the 12.7\,Jy $\rm{beam}^{-1} \rm{km} \rm{s}^{-1} = 8.8 \times 
10^{21} \rm{cm}^{-2} $ CO peak 
\citep{you05}. 
{\it Right:} 
White contours of the 
CSO/SHARC\,II 350\,$\mu$m continuum 
smoothed to $10''$ 
are overlayed on
intensity-grey scale and black contours of the 
DSS2 
 blue image of
NGC\,83.
The DSS2-blue and  
submm contours 
are respectively 
50\%, 70\%, and 90\% and 
20\%, 30\%, 50\%, 70\%, and 90\% of the maximum pixel values
on the 
maps and are  
intended to display areas that may be co-spatial with the CO as shown 
in the adjacent figure.  The DSS2 image of NGC\,83 is saturated and
plotted only as a guide of the possible optical extent of this galaxy.
\label{fig:ngc83}}
\end{center}
\end{figure}

\begin{figure}[!th]
\begin{center} 
\vspace*{-30mm}
\includegraphics[angle=0,height=4.318in]{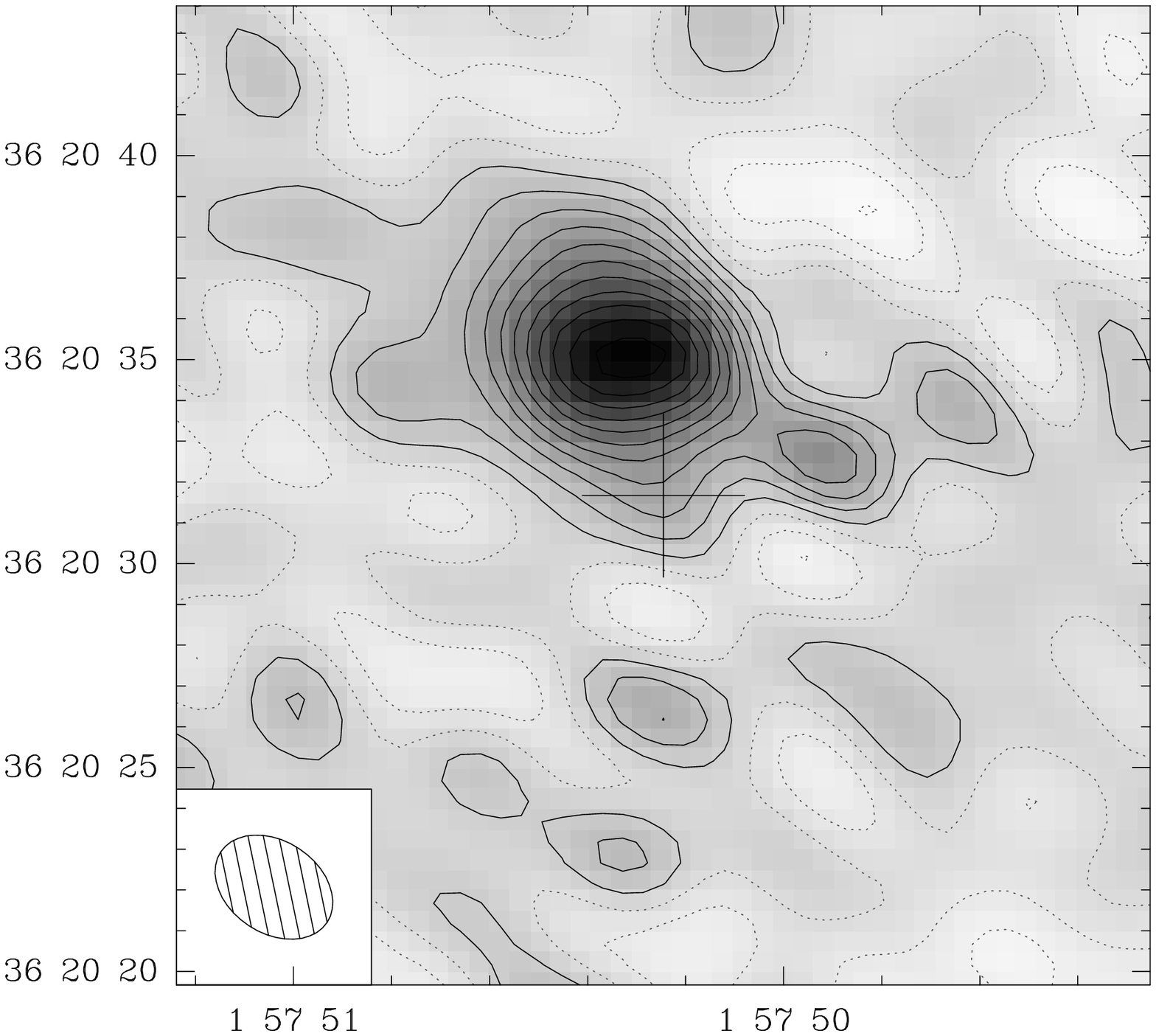}
\includegraphics[angle=0,height=2.718in]{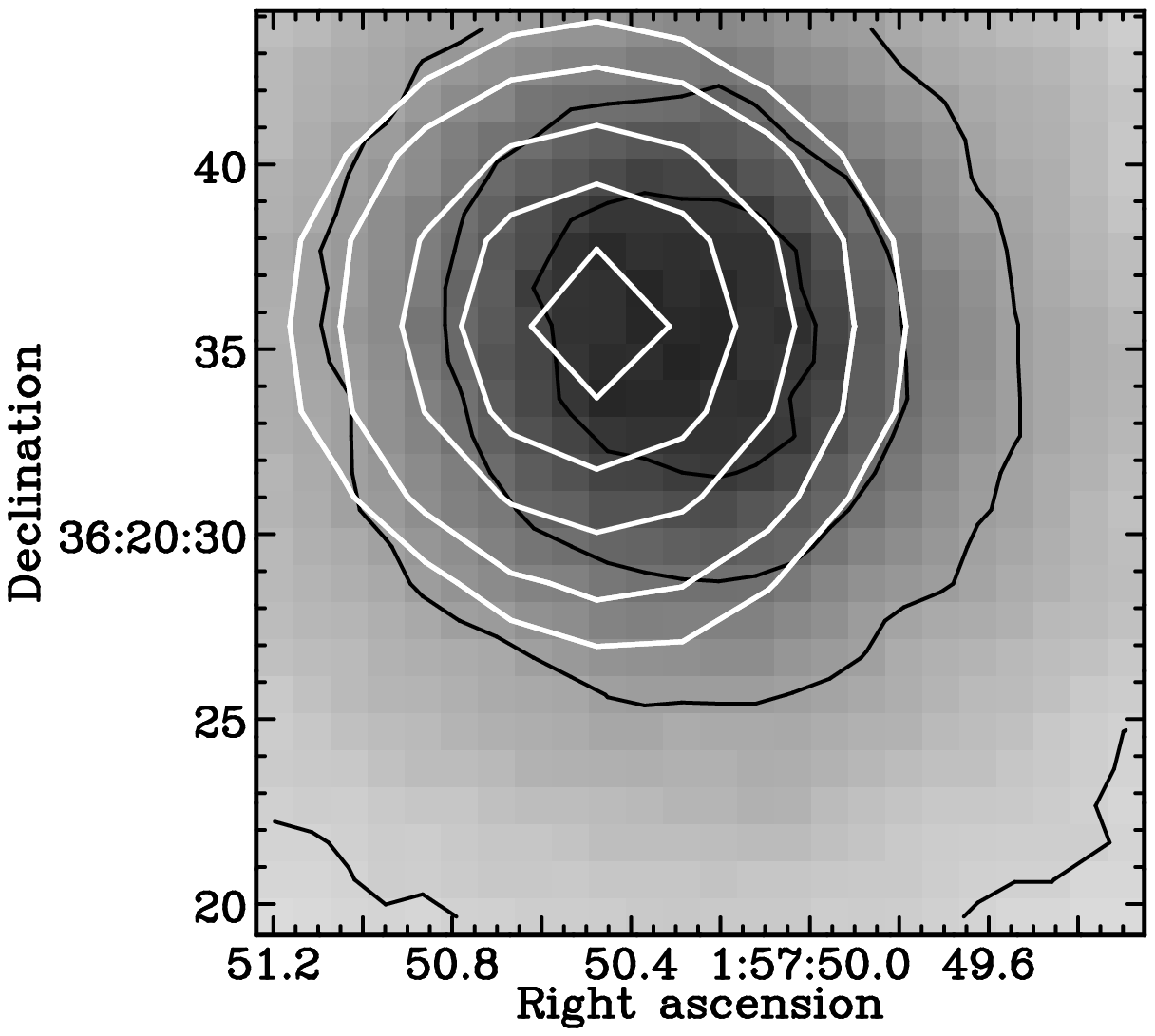}
\caption{
{\it Top Left:} 
Total integrated CO(1-0) intensity map
of NGC\,759 (grey scale and black contours) that is 
adopted from the paper by 
\citet[][]{wik97}.
The diameter of the CO, black contour level spacing, and rms noise in the map
are respectively $5''$, 0.8\,Jy $
\rm{km} \rm{s}^{-1}$, and 0.8\,Jy $ \rm{km} \rm{s}^{-1}$ 
\citet[][]{wik97}
{\it Top Right:} 
White contours of the 
CSO/SHARC\,II 350\,$\mu$m continuum 
smoothed to $10''$ 
are overlayed on
intensity-grey scale and black contours of the Digitized Sky Survey 2 (DSS2) 
 blue
image of
NGC\,759. 
The DSS2-blue and  
submm contours 
are respectively 
30\%, 
50\%, 70\%, and 90\% and 
20\%, 30\%, 50\%, 70\%, and 90\% of the maximum pixel values
on the 
maps and are  
intended to display areas that may be co-spatial with the CO as shown 
in the adjacent figure.
\label{fig:ngc759}}
\end{center}
\end{figure}

\begin{figure}[!th]
\begin{center} 
\includegraphics[angle=0,height=2.518in]{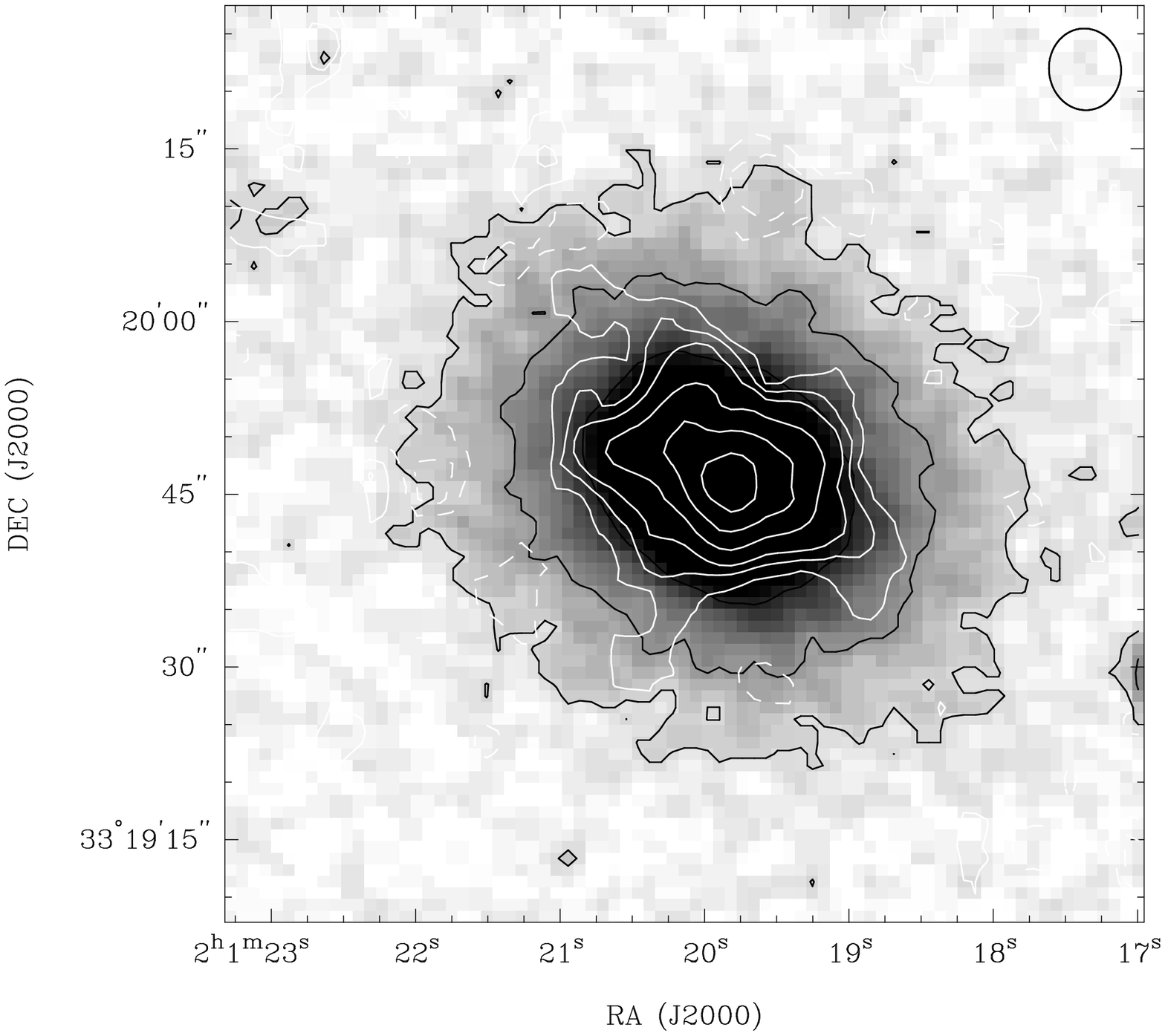}
\includegraphics[angle=0,height=2.518in]{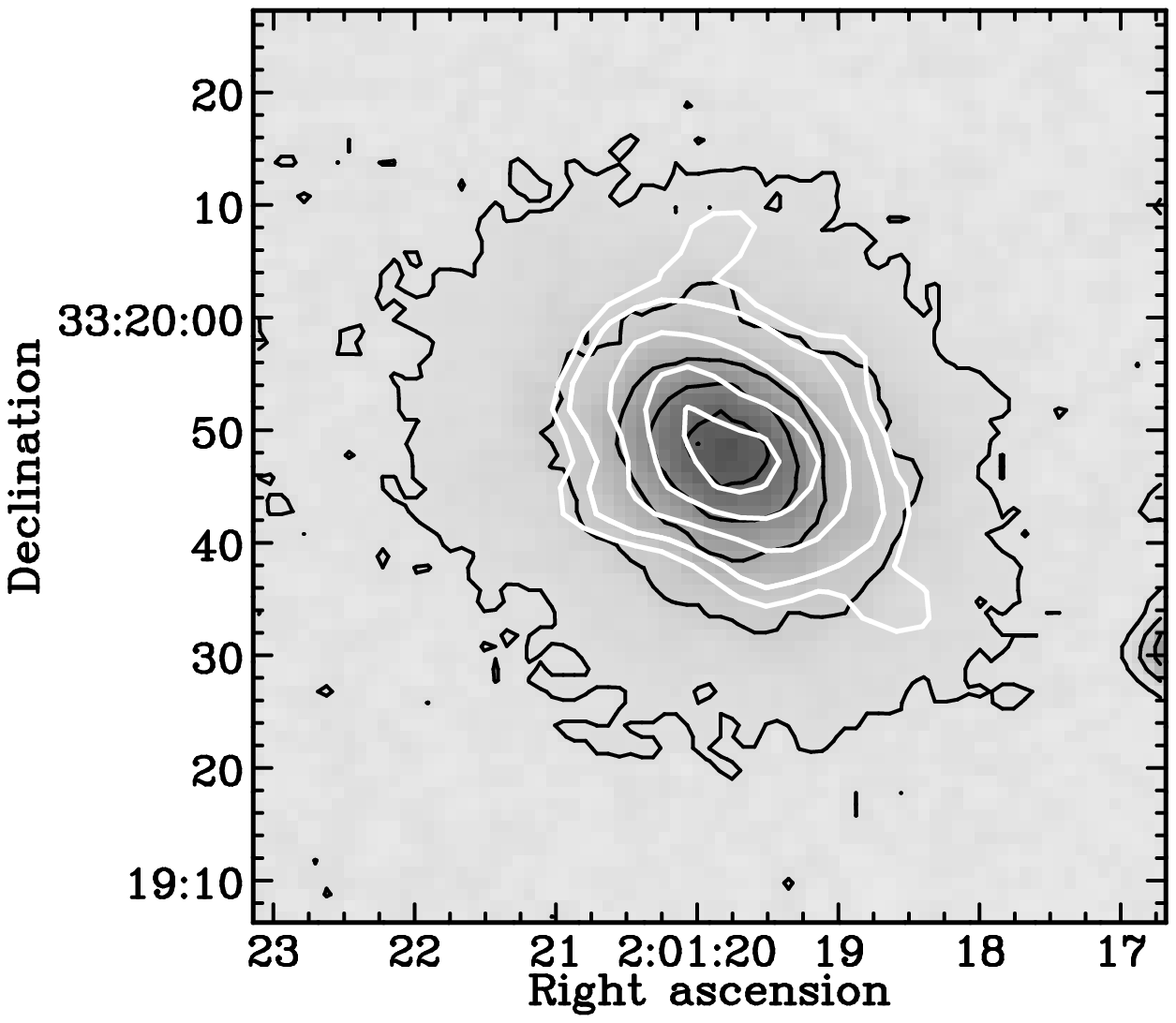}
\caption{
{\it Top Left:} 
White contours of the total integrated CO(1-0) intensity map 
of UGC\,1503, with a measured 10\%-contour-level diameter of $30''$,
are overlayed on the grey scale and black contours 
from the red part of the optical Digitized Sky Survey 2 (DSS2), as presented  by 
\citet{you02}. 
The white contours are in units 
of -20\%, -10\%, 10\%, 20\%,
30\%, 50\%, 70\%, and 90\% of 
6.3\,Jy $\rm{beam}^{-1} \rm{km} \rm{s}^{-1} = 3.9 \times 
10^{21} \rm{cm}^{-2} $ CO integrated intensity 
peak
\citep{you02}.
{\it Top Right:} 
White contours of the 
CSO/SHARC\,II 350\,$\mu$m continuum 
smoothed to $10''$ 
are overlayed on
intensity-grey scale and black contours of the 
DSS2 blue image of
UGC\,1503.
The DSS2-blue and  
submm contours 
are respectively 20\%, 30\%, 50\%, 70\%, and 90\% and 
20\%, 30\%, 50\%, 70\%, and 90\% of the maximum pixel values
on the 
maps and are  
intended to display areas that may be co-spatial with the CO as shown 
in the adjacent figure.
\label{fig:ugc1503}}
\end{center}
\end{figure}

\begin{figure}[!th]
\vspace*{-0.3cm}
\begin{center} 
\includegraphics[angle=0,height=2.518in]{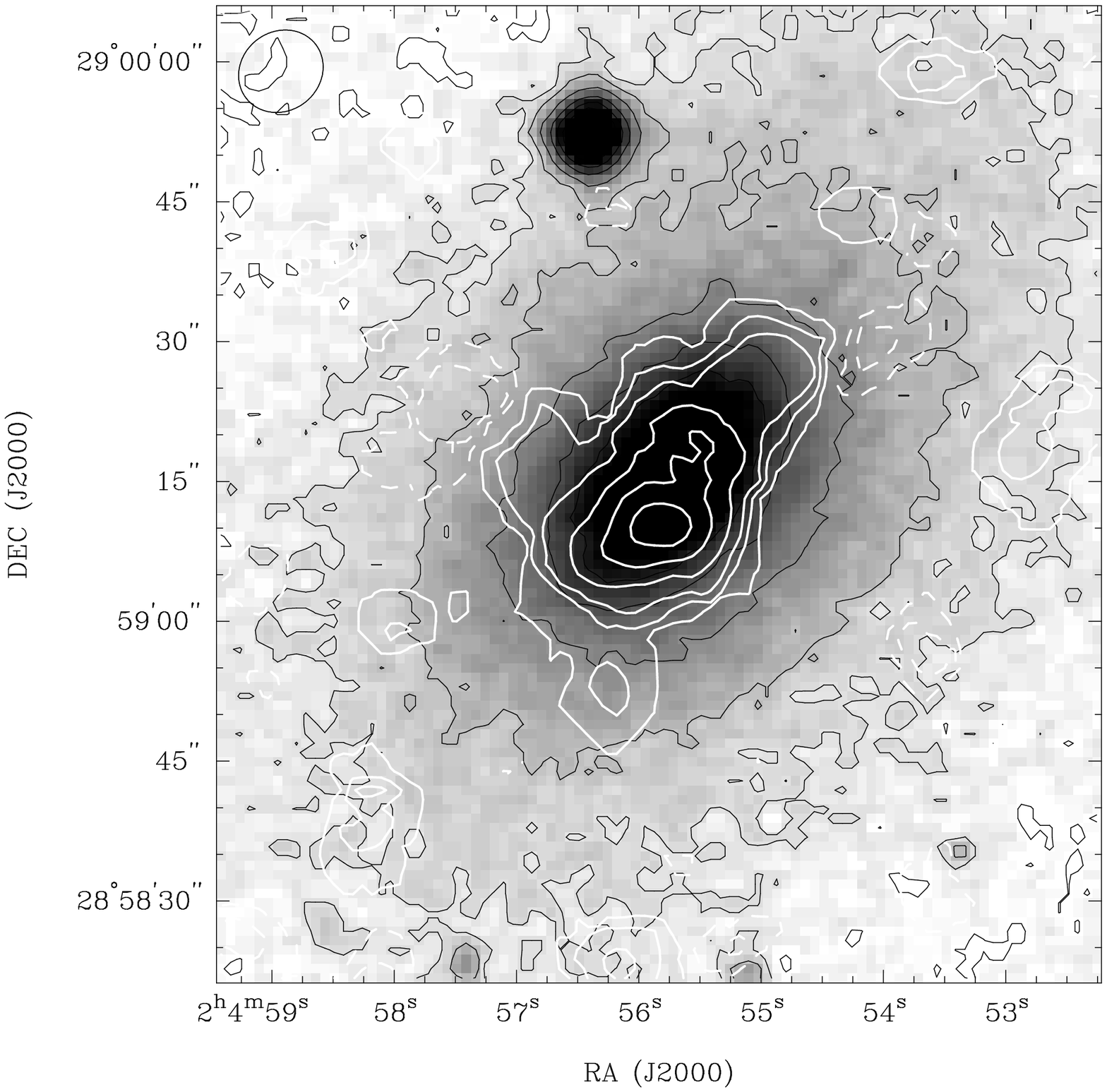}
\includegraphics[angle=0,height=2.518in]{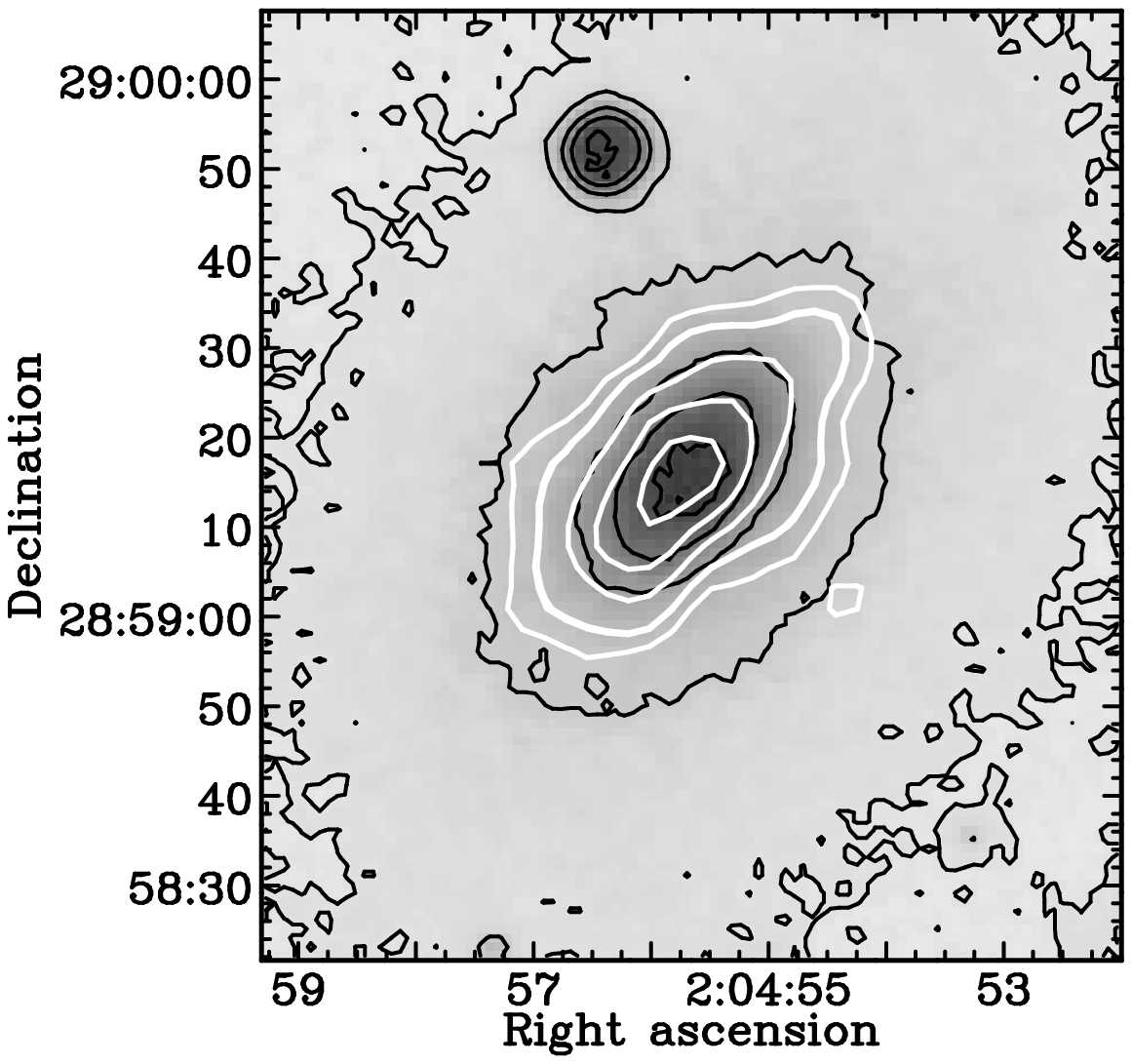}
\caption{
{\it Top Left:} 
White contours of the total integrated CO(1-0) intensity map 
of NGC\,807, with a measured 10\%-contour-level diameter of $40''$,
are overlayed on the grey scale and black contours 
from the 
red 
part of the optical Digitized Sky Survey 2 (DSS2), as presented  by 
\citet{you02}. 
The white contours are in units 
of -20\%, -10\%, 10\%, 20\%,
30\%, 50\%, 70\%, and 90\% of 
7.6\,Jy $\rm{beam}^{-1} \rm{km} \rm{s}^{-1} = 2.6 \times 
10^{22} \rm{cm}^{-2} $ 
CO peak 
\citep{you02}.
{\it Top Right:} 
White contours of the 
CSO/SHARC\,II 350\,$\mu$m continuum 
smoothed to $10''$ are overlayed on
intensity-grey scale and black contours of the 
DSS2 blue image of
NGC\,807.
The DSS2-blue and  
submm contours 
are respectively 20\%, 30\%, 50\%, 70\%, and 90\% and 
20\%, 30\%, 50\%, 70\%, and 90\% of the maximum pixel values
on the 
maps and are  
intended to display areas that may be co-spatial with the CO as shown 
in the adjacent figure.
\label{fig:ngc807}}
\end{center}
\end{figure}

The SHARC\,II contour maps are shown in 
Figures~\ref{fig:ngc3656} to~\ref{fig:ngc4476}    
and are the
result of about 3.0\,hours
of observing on NGC\,83, 7.4\,hours
on NGC\,807, 6.3\,hours 
on UCG\,1503, 2.3\,hours
on NGC\,759, 1.9\,hours 
on NGC\,3656, 2.4\,hours 
on NCG\,4476, and 1.3\,hours
on 
NGC\,5666.  
The data were analyzed 
using the SHARC\,II reduction software SHARCSOLVE, which iteratively solves
for the source image, atmospheric background, and detector gains and
offsets.  During the iterations, a constraint was applied in the map
domain: outside a diameter of $37''$ or $46''$, the source image was
assumed to be zero.  The 
$46''$ diameter was applied to the
maps of NGC\,807 and
UGC\,1503, and the 
$37''$ diameter applied to those of the other sources.
The positional accuracies
of the galaxies are based on the
pointing observations of the calibration sources.
  However, to optimize the effective resolution,
the coadded images are composed of hour-long sub-images whose
centroids have been aligned.  This relative alignment was performed
for all objects except NGC\,807 and UGC\,1503, which were too faint and
extended to have well-defined centroids after one-hour integrations.
The 350\,$\mu$m fluxes 
of these maps are given in Table~\ref{tab:iras}.  The integrated fluxes
were obtained in a 
$46''$ aperture for NGC\,807 and
UGC\,1503 and a 
$37''$ aperture for all the other target objects.

\section{Submm Morphology versus Observations from other Wavebands}\label{sec:submm_morph}  

The goal of this submm 
wavelength
study 
is to directly map
the emission from
cold dust that may be co-spatial with the optical dust features 
and/or the CO gas-disks 
presented by \citet[][]{wik97}
 and
\citet[][]{you02, you05}
 for the sample galaxies, as well as any emission from dust not detected
by the optical and CO studies.  The presented submm mapping is
sensitive to continuum emission from optically-thin dust 
and, thus, the observations
will detect emission from
dust that may be deep or in the 
far-side of the galaxies, and thus
not easily
visible in the optical images.  The detected submm
distribution and fluxes
are analysed to assess their galactic associations and the possible
mechanisms heating the dust; 
therefore, the submm mapping spatially and physically probes the 
dust's origin, in complement with
observations at other wavelengths.  

Below,
the distributions of the mapped submm emission are compared
to those of galactic components observed at other wavelengths. 
Later in
Section~\ref{sec:sfr_es}, the observed submm fluxes are used to calculate
and analyse the heating of dust by star-formation and other possible
submm-emitting, galactic sources.



\begin{figure}[!h]
\vspace*{-0.2cm}
\begin{center} 
\includegraphics[angle=0,height=2.518in]{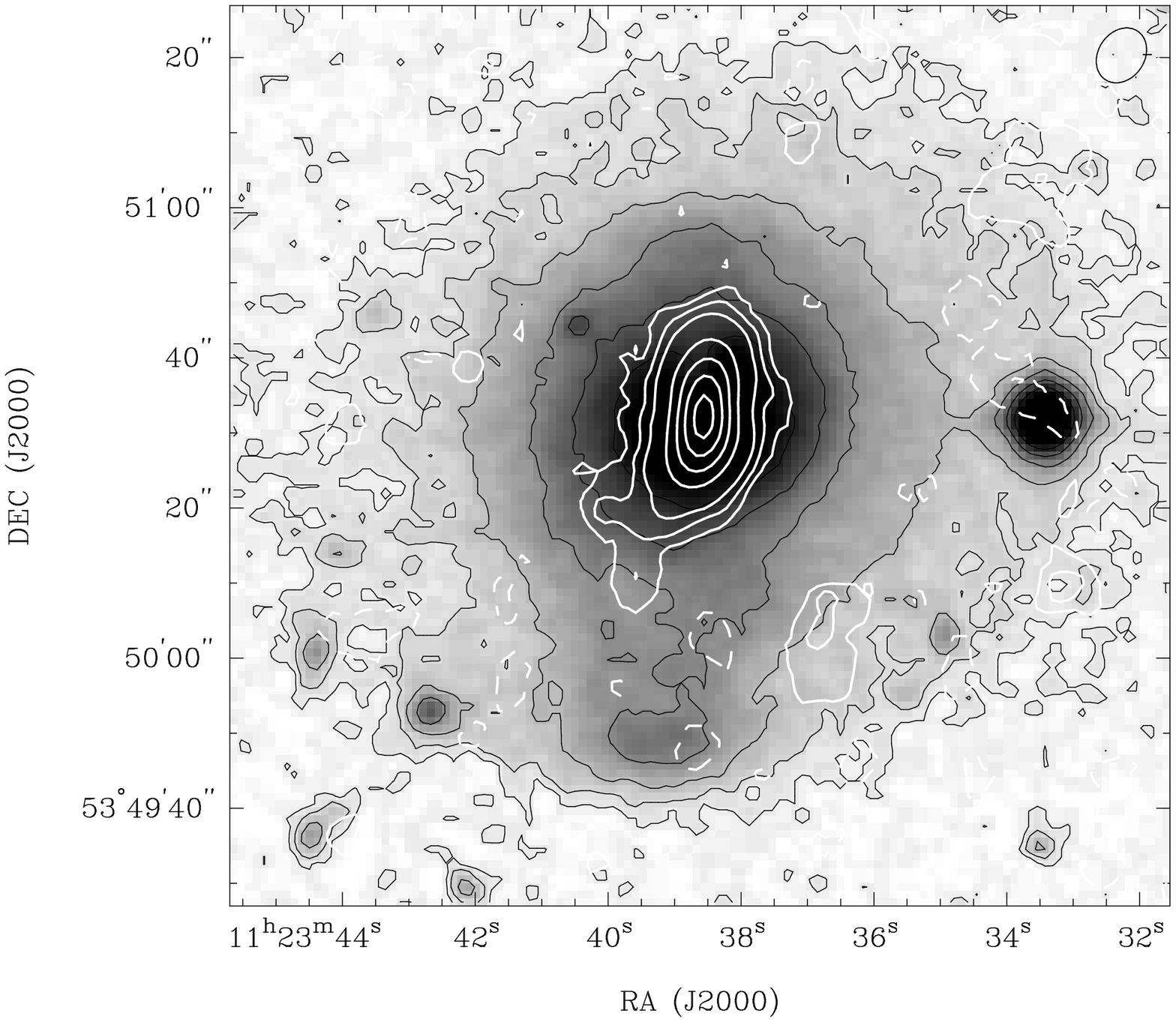}
\includegraphics[angle=0,height=2.518in]{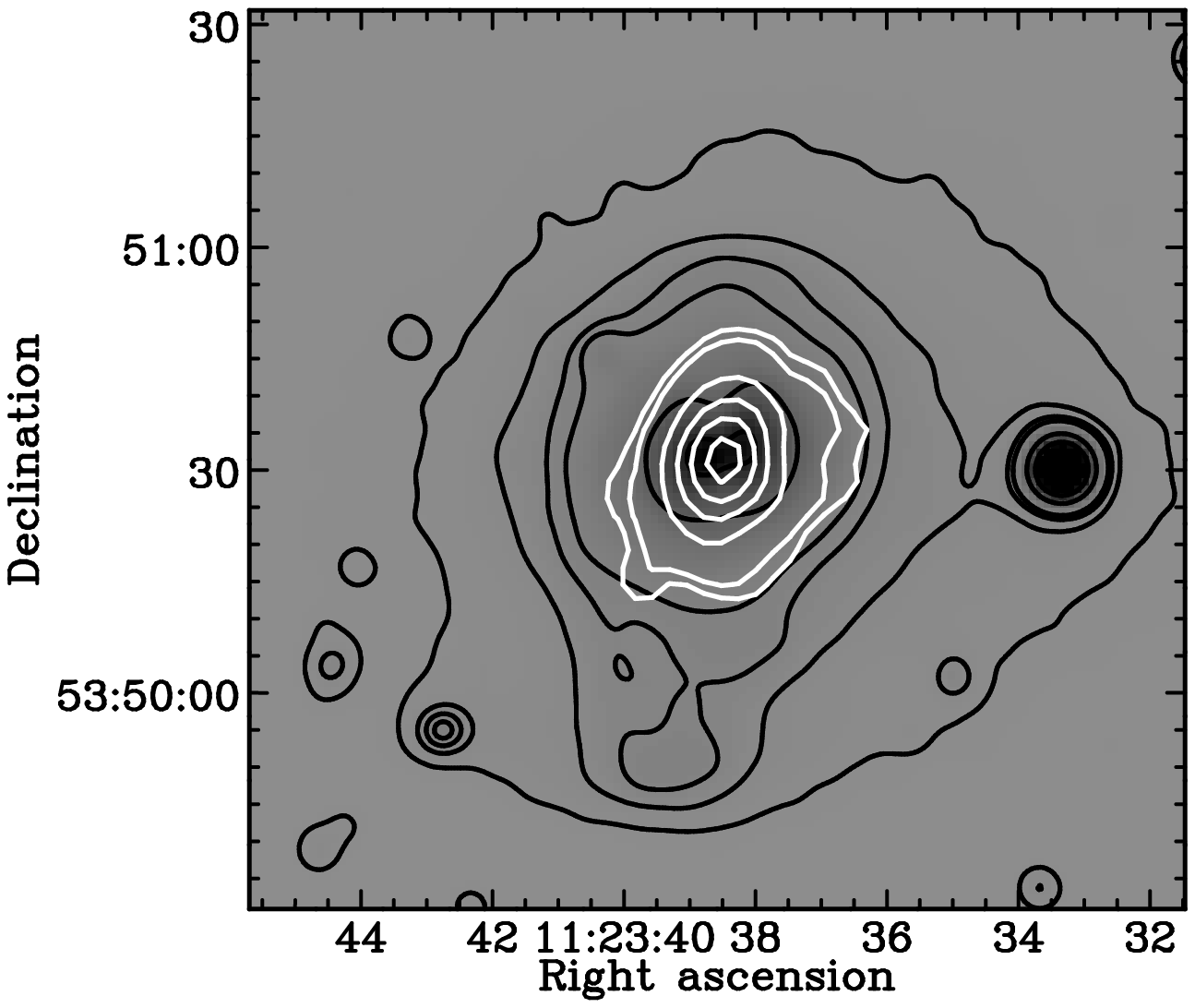}
\includegraphics[angle=0,height=2.618in]{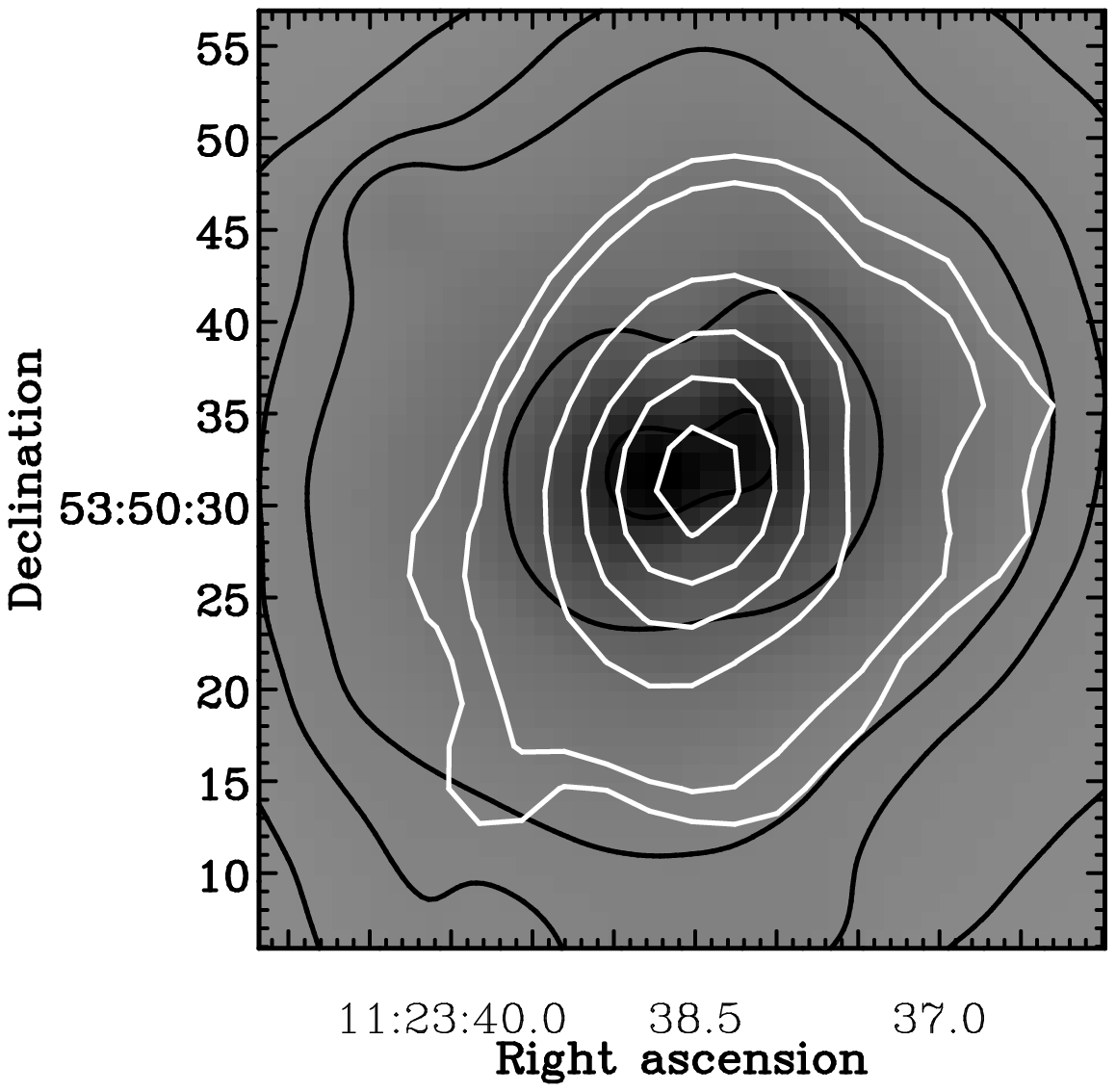}
\includegraphics[angle=0,height=2.618in]{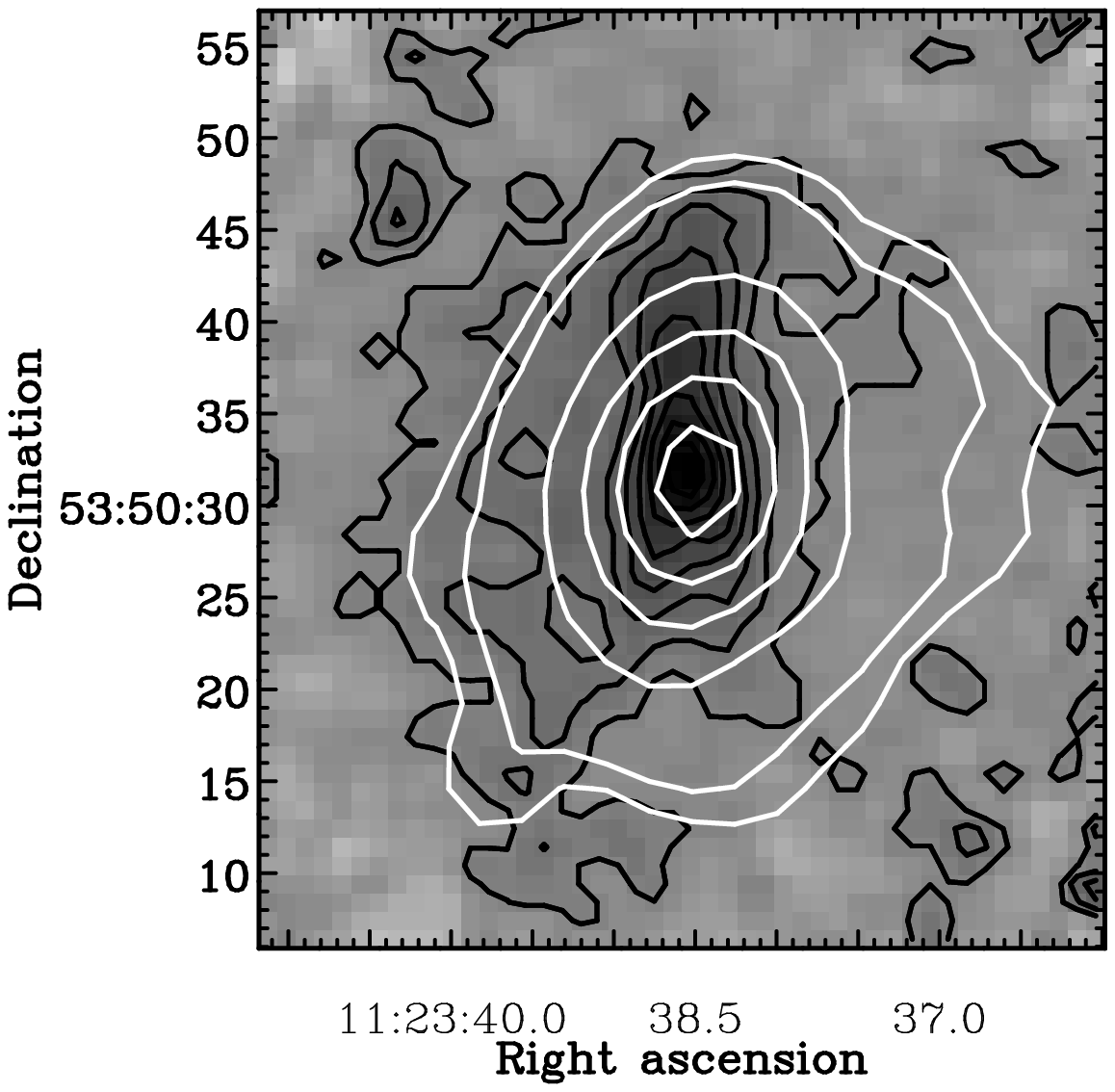}
\vspace*{-0.3cm}
\caption{
{\it Top Left:} 
White contours of the total integrated CO(1-0) intensity map 
of NGC\,3656, with a measured 10\%-contour-level diameter of $34''$,
are overlayed on the grey scale and black contours from the 
red part of the optical Digitized Sky Survey 2 (DSS2), as presented  by 
\citet{you02}. 
The white contours are in units of -5\%, -2\%, 2\%, 5\%, 10\%, 20\%,
30\%, 50\%, 70\%, and 90\% of 
81.1\,Jy $\rm{beam}^{-1} \rm{km} \rm{s}^{-1} = 4.7 \times 
10^{22} \rm{cm}^{-2} $
CO peak 
\citep{you02}.
{\it Top Right:} 
White contours of the 
CSO/SHARC\,II 350\,$\mu$m continuum 
smoothed to $10''$ are overlayed on
intensity-grey scale and black contours of the 
Sloan Digital Sky Survey (SDSS) i-band 
image of
NGC\,3656 smoothed to $10''$.
The SDSS i-band
and submm contours 
are respectively 
$\sim 57\%$, 58\%, 59\%,
60\%, 70\%, and 90\% 
and 10\%, 
30\%, 50\%, 70\%, and 90\% of the maximum pixel values
on the 
maps and are  
intended to display areas that may be co-spatial with the CO as shown 
in the adjacent figure.
{\it Bottom left:} A zoom-in of the {\it top-right} figure.
{\it Bottom right:} 
White contours of the 
CSO/SHARC\,II 350\,$\mu$m continuum 
smoothed to $\sim 10''$ are overlayed on
grey scale and black contours of the 
SDSS i- minus g-band images 
smoothed to $2''$.
The 
SDSS contours are on a linear scale,
and the submm contours are as above. 
In the 
SDSS i- minus g-band maps, darker pixels have
 redder colors.
\label{fig:ngc3656}}
\end{center}
\vspace*{-0.3cm}
\end{figure}

\begin{figure}[!h]
\vspace*{-0.2cm}
\begin{center} 
\includegraphics[angle=0,height=2.518in]{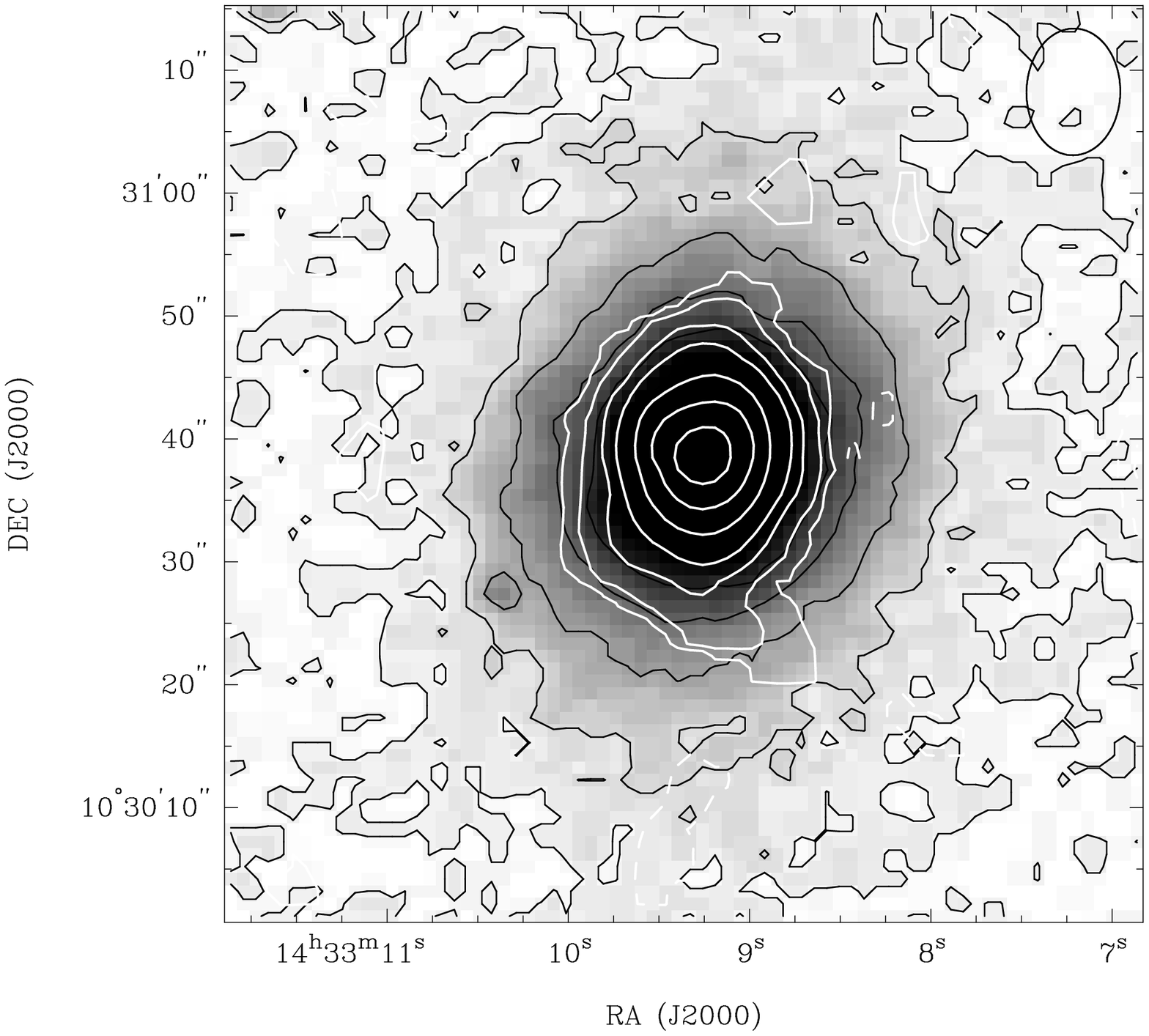}
\includegraphics[angle=0,height=2.518in]{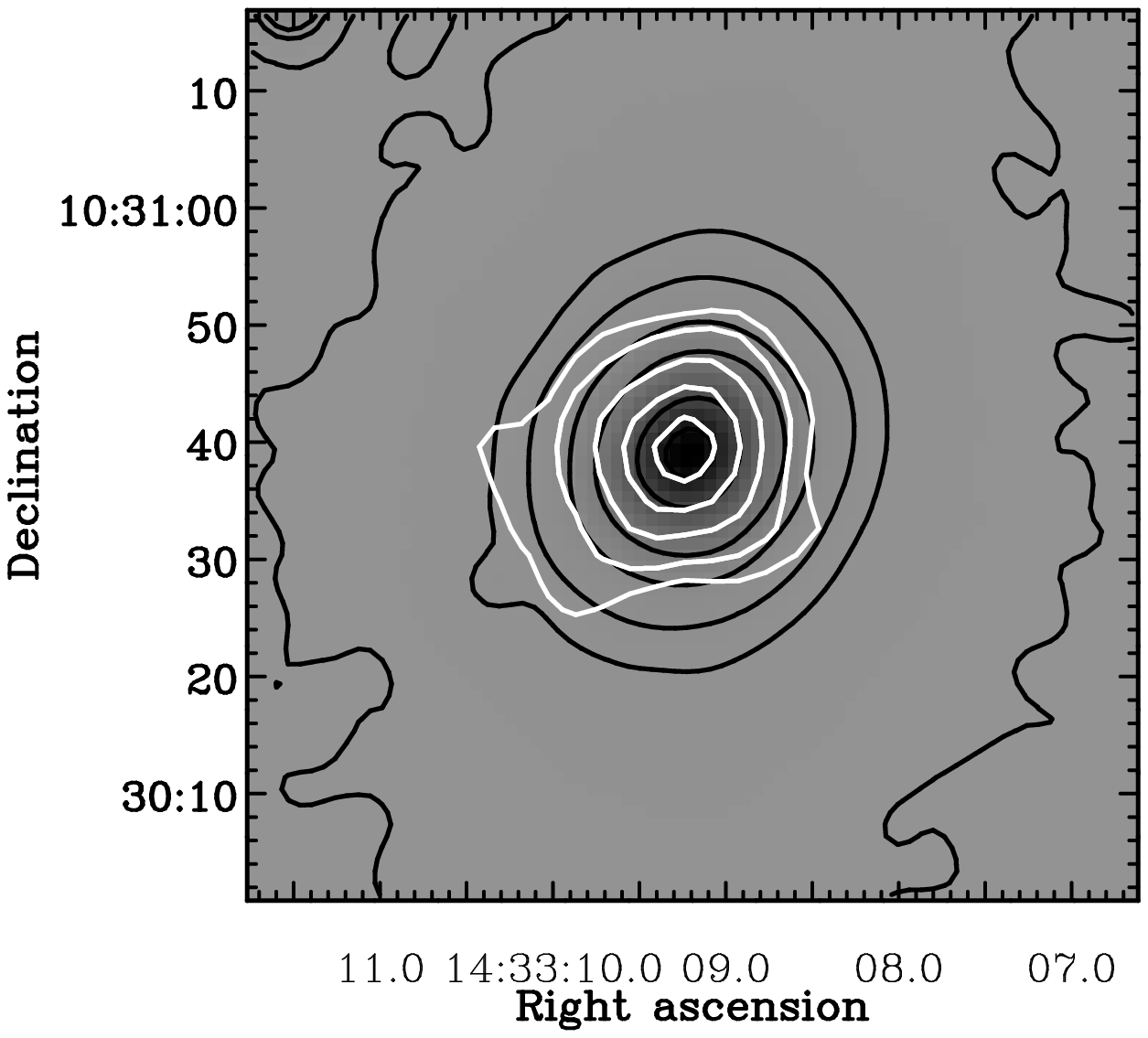}
\includegraphics[angle=0,height=2.518in]{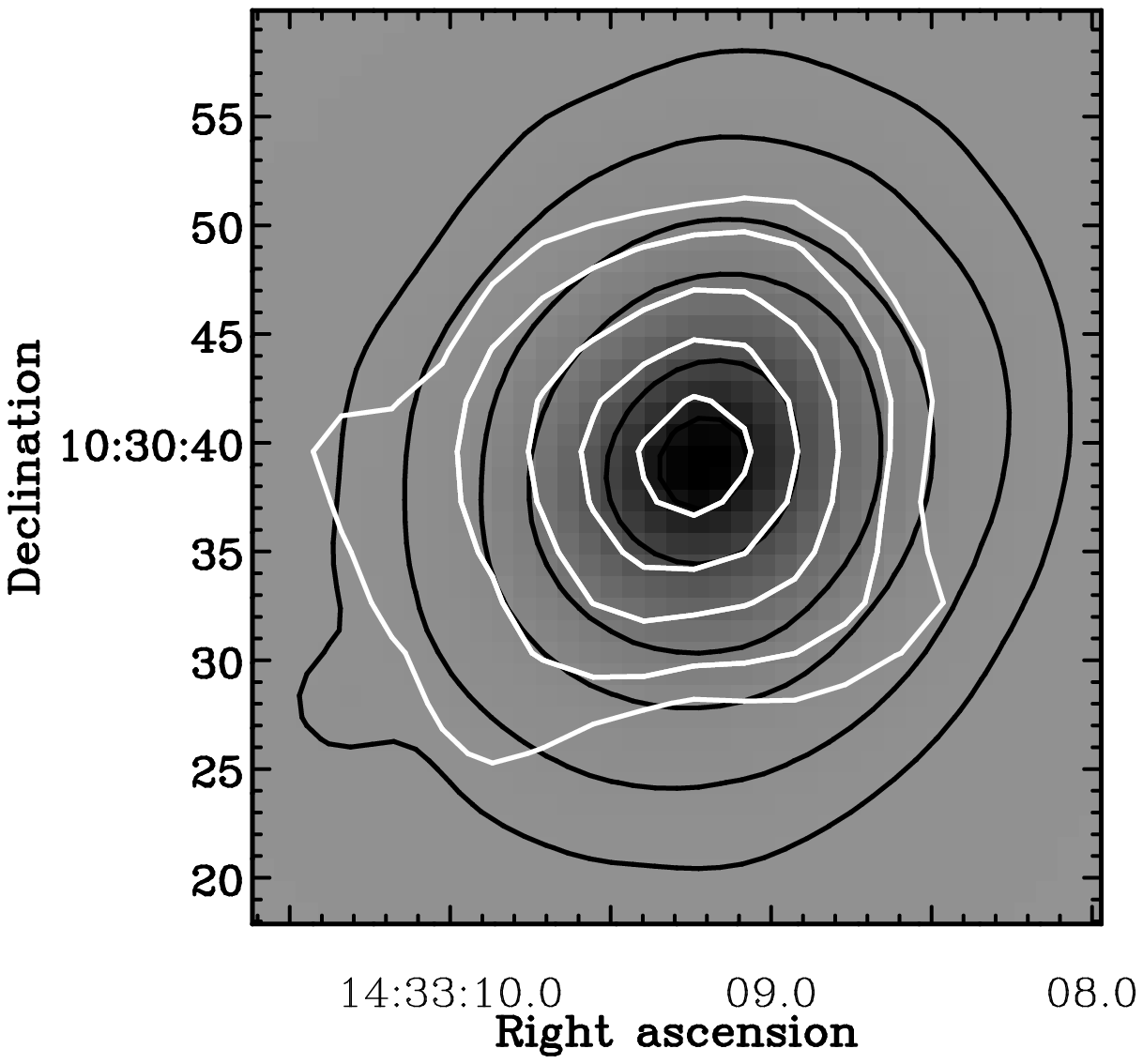}
\includegraphics[angle=0,height=2.518in]{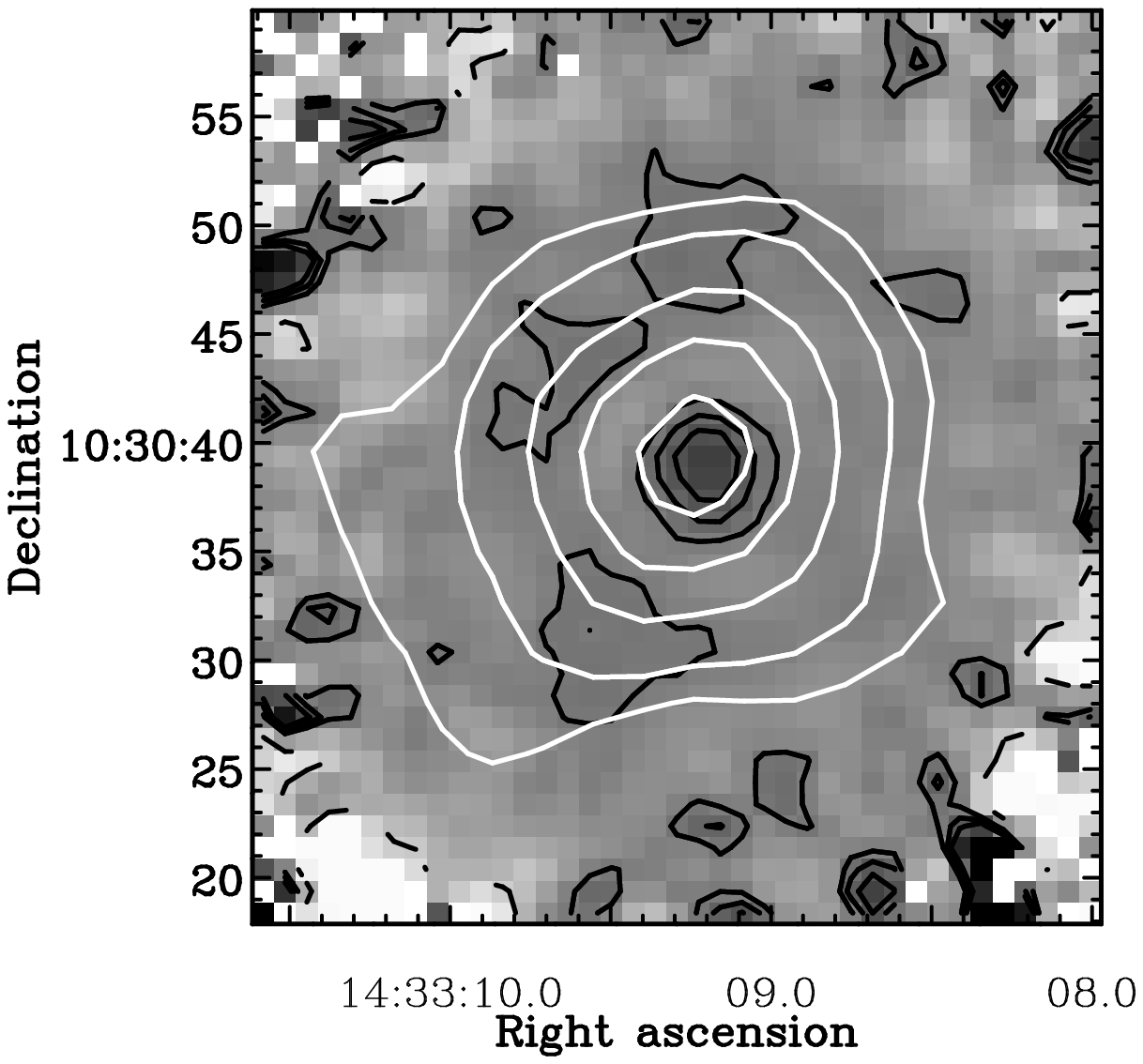}
\vspace*{-0.3cm}
\caption{
{\it Top Left:} 
White contours of the total integrated CO(1-0) intensity map 
of NGC\,5666, with a measured 10\%-contour-level diameter of $26''$,
are overlayed on the grey scale and black contours from the 
red part of the optical Digitized Sky Survey 2 (DSS2), as presented  by 
\citet{you02}. 
The white contours are in units of 
-10\%, -5\%, 5\%, 10\%, 20\%,
30\%, 50\%, 70\%, and 90\% of 
21.3\,Jy $\rm{beam}^{-1} \rm{km} \rm{s}^{-1} = 7.5 \times 
10^{21} \rm{cm}^{-2} $ 
CO peak 
\citep{you02}.
{\it Top Right:} 
White contours of the 
CSO/SHARC\,II 350\,$\mu$m continuum 
smoothed to $10''$ 
are overlayed on
intensity-grey scale and black contours of the 
Sloan Digital Sky Survey (SDSS) i-band 
image of NGC\,5666
smoothed to $10''$.
The 
SDSS i-band
and  
submm contours 
are respectively 
$\sim 40\%$, 41\%, 42\%, 45\%, 50\%, 70\%, and 90\% and 
20\%, 30\%, 50\%, 70\%, and 90\% of the maximum pixel values
on the 
maps and are  
intended to display areas that may be co-spatial with the CO as shown 
in the adjacent figure.
{\it Bottom left:} A zoom-in of the {\it top-right} plot.
{\it Bottom right:} White contours of the 
CSO/SHARC\,II 350\,$\mu$m continuum 
smoothed to $\sim 10''$ are overlayed on
grey scale and black contours of the 
SDSS i- minus g-band images 
smoothed to $2''$. 
The 
SDSS contours are on a linear scale,
and the submm contours are as above. 
In the 
SDSS i- minus g-band maps, darker pixels have
 redder colors.
\label{fig:ngc5666}}
\end{center}
\vspace*{-0.3cm}
\end{figure}

\begin{figure}[!h]
\vspace*{-0.2cm}
\begin{center} 
\includegraphics[angle=0,height=2.518in]{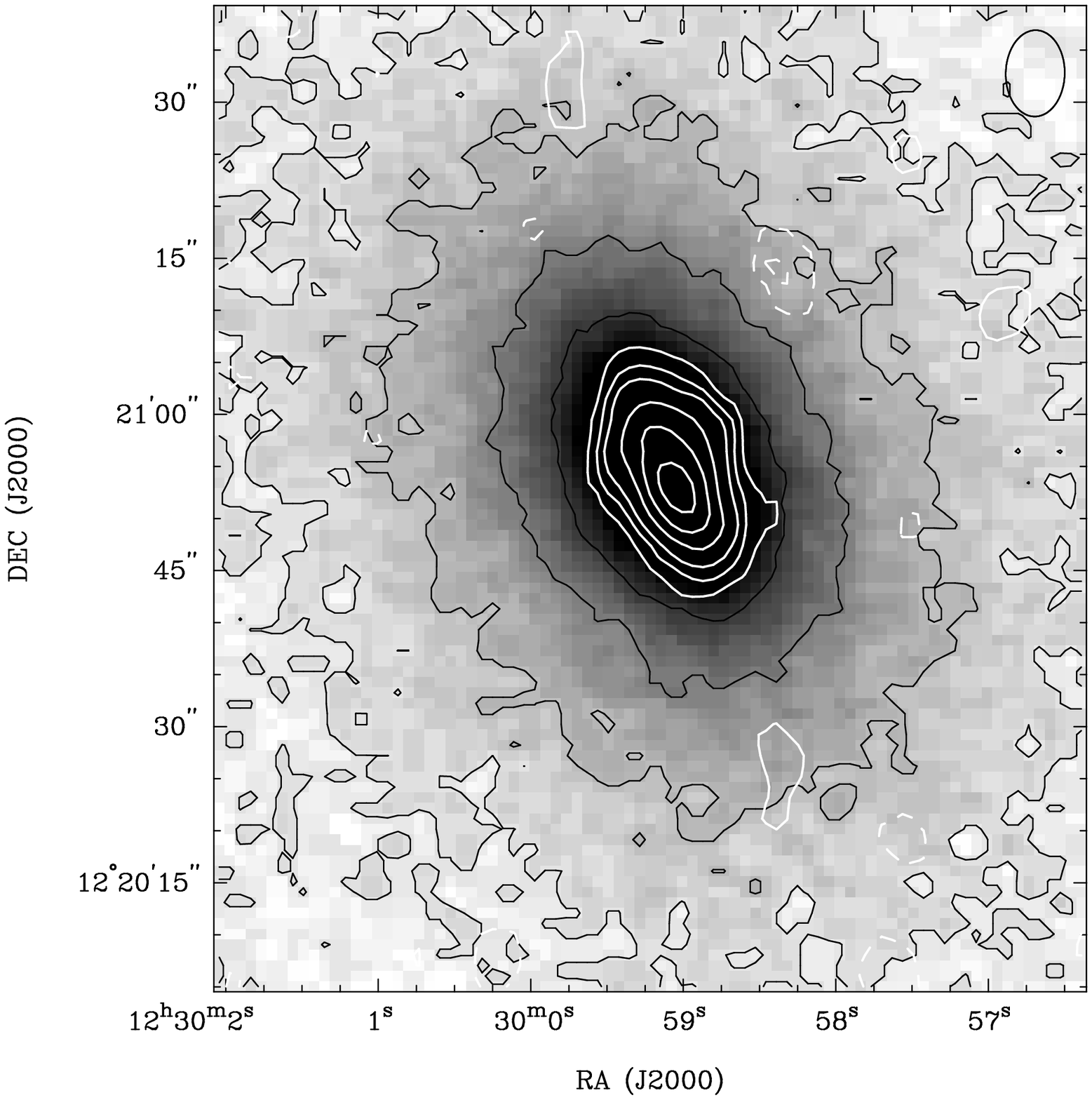}
\includegraphics[angle=0,height=2.518in]{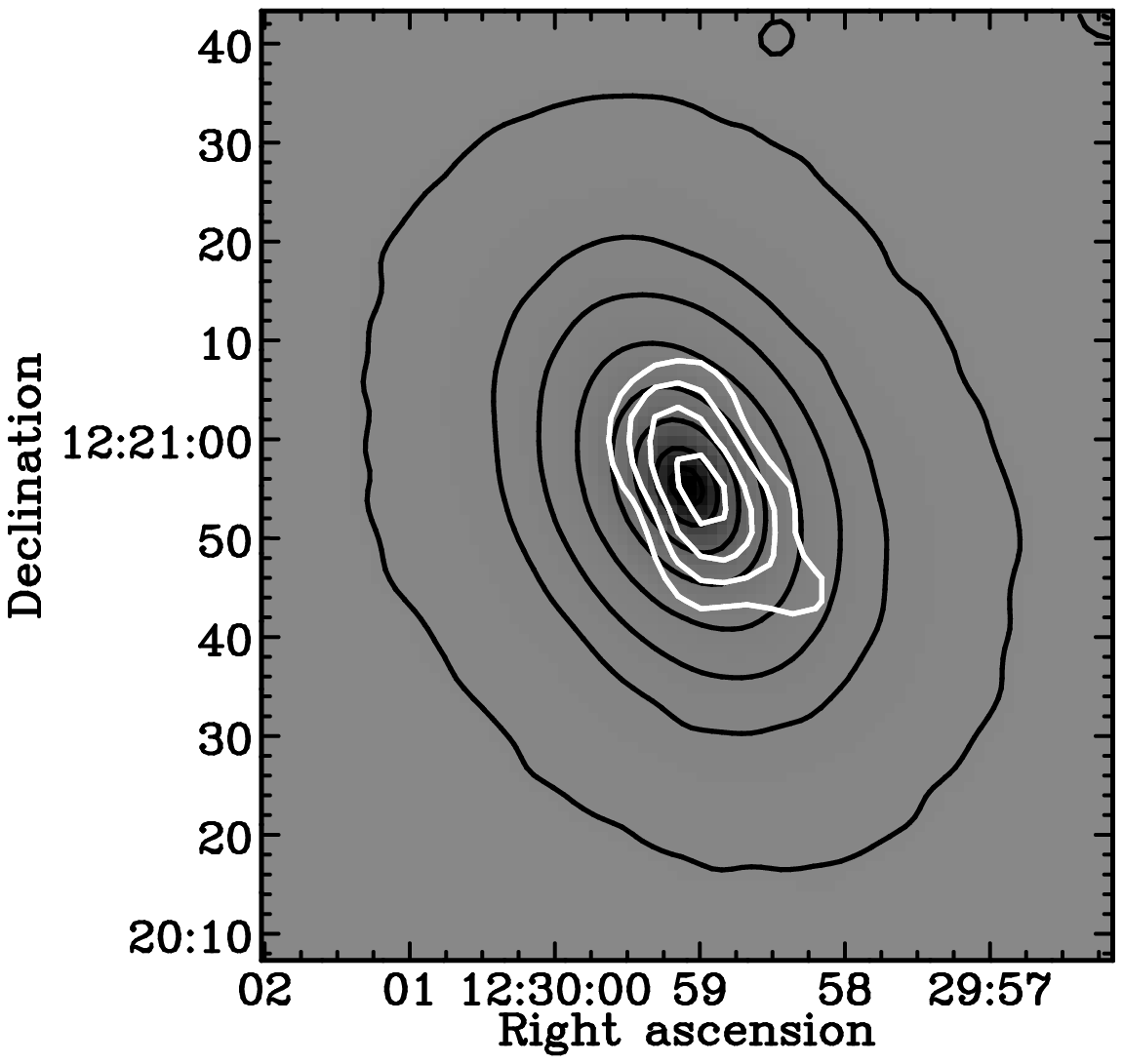}
\includegraphics[angle=0,height=2.518in]{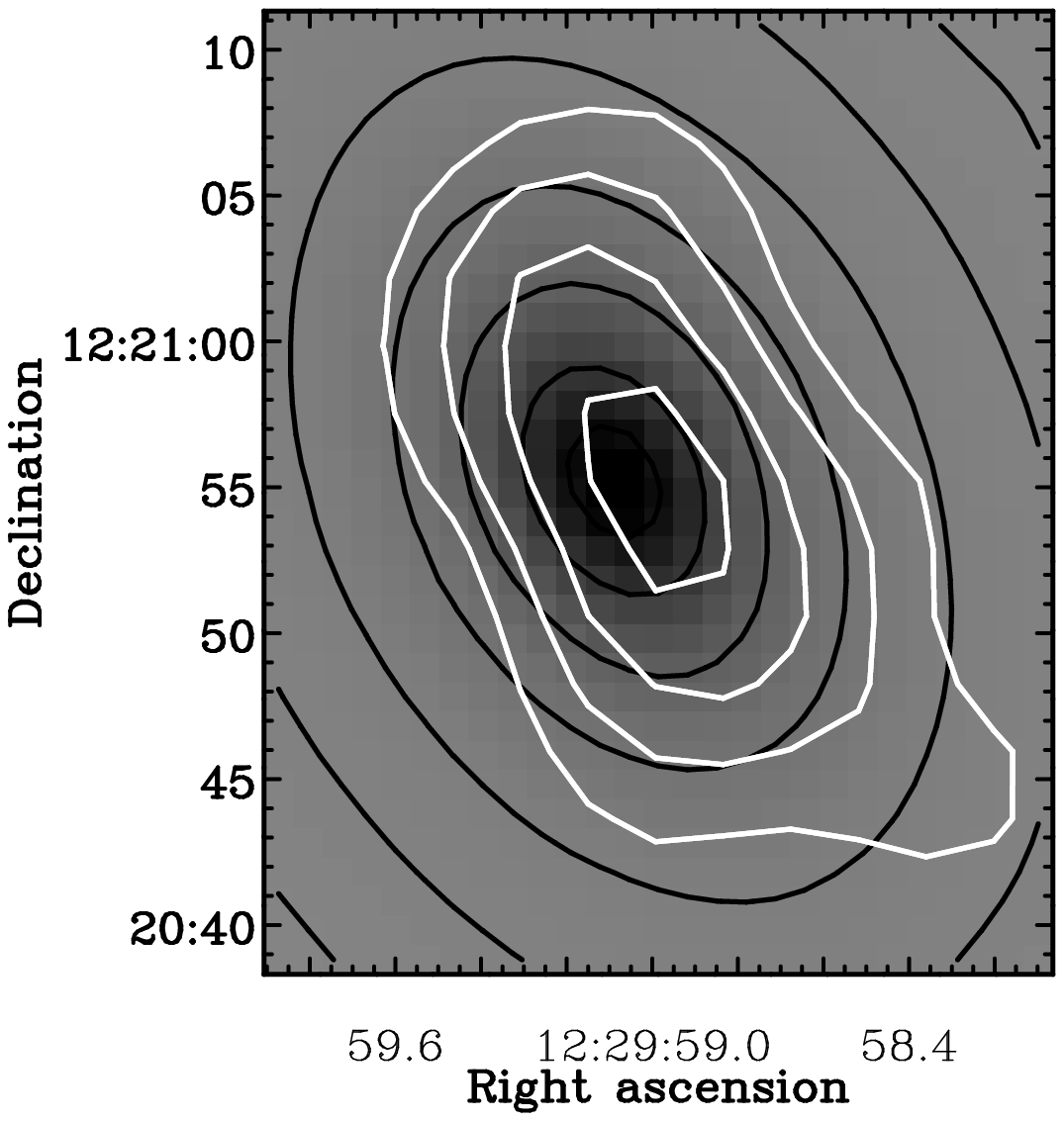}
\includegraphics[angle=0,height=2.518in]{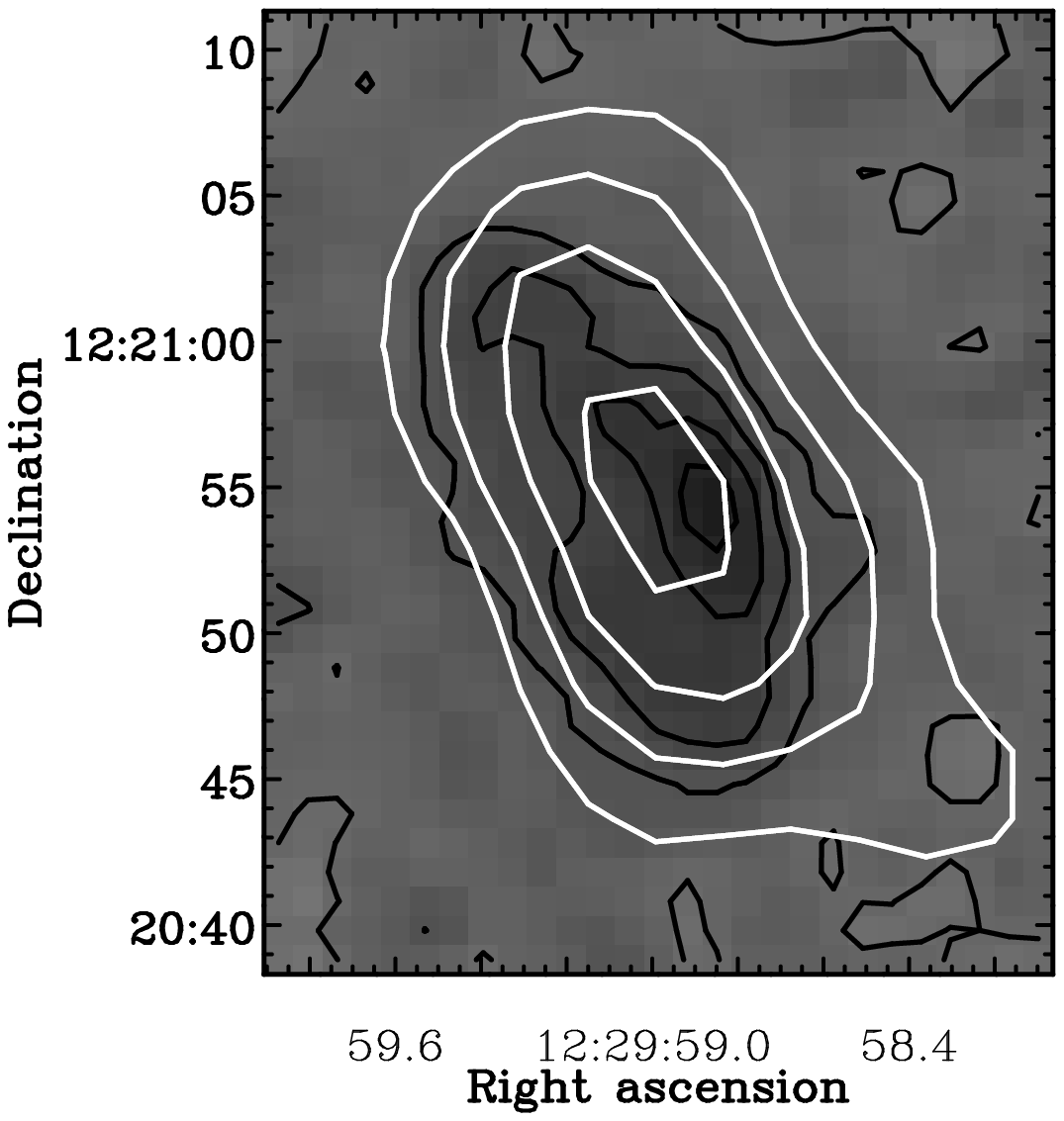}
\vspace*{-0.3cm}
\caption{
{\it Top Left:} 
White contours of the total integrated CO(1-0) intensity map 
of NGC\,4476, with a measured 10\%-contour-level diameter of $27''$,
are overlayed on the grey scale and black contours from the 
red part of the optical Digitized Sky Survey 2 (DSS2), as presented  by 
\citet{you02}. 
The white contours are in units of 
-20\%, -10\%, 10\%, 20\%,
30\%, 50\%, 70\%, and 90\% of 
12.4\,Jy $\rm{beam}^{-1} \rm{km} \rm{s}^{-1} = 7.3 \times 
10^{21} \rm{cm}^{-2} $
CO peak 
\citep{you02}.
{\it Top Right:} 
White contours of the 
CSO/SHARC\,II 350\,$\mu$m continuum 
smoothed to $10''$ are overlayed on
intensity-grey scale and black contours of the 
Sloan Digital Sky Survey (SDSS) i-band 
image of NGC\,4476
smoothed to $10''$.
The 
SDSS i-band and  
submm contours 
are respectively 
$\sim 31\%$, 32\%, 33\%, 35\%, 40\%, 50\%, 70\%, and 90\% and 
30\%, 50\%, 70\%, and 90\% of the maximum pixel values
on the 
maps and are  
intended to display areas that may be co-spatial with the CO as shown 
in the adjacent figure.
{\it Bottom left:} A zoom in of the {\it top-rright} figure. 
{\it Bottom right:} 
White contours of the 
CSO/SHARC\,II 350\,$\mu$m continuum 
smoothed to $\sim 10''$ are overlayed on
grey scale and black contours of the 
SDSS i- minus g-band images 
smoothed to $2''$.
The 
SDSS contours are on a linear scale,
and the submm contours are as above. 
In the 
SDSS i- minus g-band maps, darker pixels have
 redder colors.
\label{fig:ngc4476}}
\end{center}
\vspace*{-0.3cm}
\end{figure}

Figures~\ref{fig:ngc3656} to~\ref{fig:ngc4476}      
show contours of the detected
350\,$\mu$m dust continuum  for the
sample
and the total integrated CO(1-0) 
intensity maps from 
\citet[][]{you02, you05}
 and 
\citet[][]{wik97},
overlayed on the optical Sloan Digital Sky Survey (SDSS) maps, and where there are no SDSS images available on the 
Digitized Sky Survey 2 (DSS2) images.  
The SDSS images have more reliable photometry than the DSS2 images; and,
the SDSS images are plotted convolved to a $10''$ beam
similar to that of the CSO dust emission images, while the DSS2 images are
plotted at their 
full resolution. 
Figures~\ref{fig:ngc3656}, ~\ref{fig:ngc5666}, and ~\ref{fig:ngc4476} 
also show
SDSS i- minus g-band color or extinction maps and their linear
contours in an attempt to show some of the dust's 
geometry.
For a smooth presentation, these color maps are shown 
convolved to a $2''$ beam. 

The SDSS and DSS2 intensity contours are percentages of the respective peaks in the maps and are
intended to display areas that may be co-spatial at the different
wavelengths.  The  50\%-contour-level diameters determined by eye 
from these plots
after deconvolution are listed in Table~\ref{tab:cso_2}. For
NGC\,3656, the optical 50\%-contour-level diameter is well off the maps
flotted in Figure~\ref{fig:ngc3656}, and therefore only the optical
70\%-contour-level diameter is listed in Table~\ref{tab:cso_2}.   
These tabulated image sizes were derived after deconvolving beamsizes
of $7''$ for the CO images (except for the size of NGC\,759, where a
$3''$ beam was used) and $10''$ for the submm and SDSS images.  For the
submm and SDSS images, the 
diameter
measurements done by eye were 
compared with FWHM diameters from 2D gaussian
fits assuming elliptical shapes and agreed to within about $\sim 5$\%.  For
the DSS2 images, the diameters listed are from the full resolution
maps.  
The relative percentage contours of the 
optical maps will
be affected by the distribution of dust in these galaxies, 
therefore the optical intensity contours on 
those maps are not a simple reflection
of the stellar distributions 
as shown on those maps and listed in Table~\ref{tab:cso_2}; but rather,
they show the stars plus their
attentuation by internal galactic dust of varying distributions, and for the
DSS2 images may also show saturation from the original photographic images.

The 
350\,$\mu$m results 
show that the detected
 dust continuum of all the 
presented 
ellipticals
generally
follows the extended CO emission 
that has been previously detected in these galaxies.  This is
therefore consistent with emission
from cold dust
that occurs in 
dense, gas-rich star-formation regions, as is commonly seen in mapping
observations of the Milky Way and
other star-forming galaxies 
\citep[e.g.,][]{sea01}
and confirms
CO-FIR correlations of unresolved data from elliptical galaxies that have
suggested 
that cold gas and dust co-exist 
in ellipticals 
\citep[e.g.,][]{lee91}.
Individual galaxies are discussed in the following sub-sections.

\subsection{The Early-Age Major-Merger/Accretion Elliptical NGC\,3656}
\label{subsec:3656}

Resolved 
imaging 
of NGC\,3656 
by \citet[][]{bal01}
showed that H\,I
occurs in shells and
tidal tails that are thought to be signatures of a recent (or {early-age}), major
merger of this galaxy.  The H\,I is more extended than the galaxy's
main optical component.
Sensitive CCD optical images of the dust 
features in 
NGC\,3656 
were studied by 
\citet[][]{bal97}
 and 
\citet[][]{rot06},
who 
suggest that the dust is a remnant of a dust-rich merger that involved this
galaxy's primary optical elliptical component.

Figure~\ref{fig:ngc3656} shows that large-scale, 
350\,$\mu$m-dust continuum spurs in the 
southeast
of 
NGC\,3656 
generally follow the 
CO-emission and optical spurs respectively presented by 
\citet[][]{you02}
 and
\citet[][]{bal97}.
 The 
350-$\mu$m continuum spurs also follow the optical spurs shown by 
\citet[][]{bal97} in the
west
of this early-merger galaxy.
These spurs may be large-scale 
features of a warped-disk  of dust that has been reported in the center
of NGC\,3656 \citep[e.g.][]{bal97} and described further below. 

Figure~\ref{fig:ngc3656} 
also shows 
that 
the extinction contours of 
NGC\,3656 are 
very
asymmetric, depicting
strong extinction in the galactic 
east
and 
dense north-south central dust
distribution.
%
The 
dust in 
NGC\,3656
is probably
in a thin, warped disk that provides the strongest extinction where
the disk is tangent to the line of slight.  In the center, the thin disk is
therefore tangent to the line of sight and thus provides the strong optical extinction
and submm emission seen there, while it is more face-on or in the
background in the western part of the galaxy.  In the east of the galaxy, the
disk is either more 
edge-on or in the foreground,
and thus exhibits the moderate
optical extinction seen there (see Figure~\ref{fig:ngc3656}).  


A thin,
warped disk 
similar to that described for 
NGC\,3656 has also been 
observed in another early-merger
 galaxy Centaurus\,A 
\citep[][]{lee02, qui06},
 showing
similar extinction and submm emission features as seen in NGC\,3656.
In Centaurus\,A, it has been speculated that the disk has formed from
dust-rich material accreted from a merger that involved a gas-rich
spiral galaxy 
\citep[e.g.,][]{qui06}.
 The similarity of the warped dust-disk together with
the H\,I shells 
\citep[][]{bal01}
 detected in NGC\,3656 to those same features observed in
Centaurus\,A suggests these two galaxies have a similar evolutionary history
and that NGC\,3656 may indeed be a younger, more distant analog of
Centaurus\,A.  

\subsection{The Intermediate-Age Merger/Accretion Elliptical NGC\,5666}
\label{subsec:5666}

The H\,I in NGC\,5666 
extends well beyond the optical galaxy; 
and, it is
in circular orbits that indicate it is in dynamic equilibirum 
\citep[][]{lak87},
since
a not too recent ({or intermediate-age}) merger/accretion event that
is thought to have occured in this galaxy 
\citep[e.g.,][]{don03}.
Using the Canada France Hawaii Telescope (CFHT), \citet[][]{don03}
obtained optical CCD spectroscopy and imaging that revealed a ``spiral''
 dusty structure with tidal tails and H\,II regions 
this galaxy. The
 ``spiral'' and tidal tails are revealed after subtracting a model of the 
 bulge plus disk of the host galaxy.  

Figure~\ref{fig:ngc5666} shows that
CO emission and submm continuum contours of NGC\,5666 follow each
 other in the high-brightness region greater than the 50\% contour
level.  Outside the $10 ''$-radius the CO emission contours are more
 North-South while the submm ones (which are not detected at a high
signal-to-noise) are East-West and slightly more extended than the CO.  
In the eastern part of the galaxy,
the submm and optical SDSS i-band contour levels follow each other  
down to the 30\% level.  However, in the western part, these contour
only follow each other to the 70\% levels.  The asymmetry in the
contour correspondence 
supports the idea that the dust distribution or
extinction in NGC\,5666 is not even, but for example in a ``spiral''
structure as suggested by \citet[][]{don03}. 

Figure~\ref{fig:ngc5666} also shows a red
 extinction region within the $10''$ diameter that coincides with  the
CO emission and submm continuum
 peaks and the
 location of the ``spiral''  that was revealed in this galaxy by
\citet[][]{don03}. This extinction map 
also shows 
%
a patchy region outside $15''$-radius that may coincide with foreground
dust patches or 
the
location of the
optical tidal tails connected to the "spiral" dusty structure.


\subsection{The ``Late-Merger/Accretion'' Ellipticals NGC\,83 and NGC\,759}
\label{subsec:83_759}


NGC\,083 and NGC\,759 are elliptical galaxies with surface brightness
profiles 
that are 
fit by the ${\rm r}^{1/4}$ law
\citep[e.g.][]{you05, wik97}.
Using deep CCD $V$ and $R$ images, \citet[][]{you05} 
confirmed 
in detail that
NGC\,83 is 
classically
 fit by the ${\rm r}^{1/4}$ law 
from about radii $6''$ to $80''$ and that within about radii $6''$
it has a regular-shaped cold gaseous dust disk. NGC\,759 is also an
elliptical galaxy 
in which \citet[][]{wik97} earlier showed that there is
a compact, star-forming, central cold gaseous disk 


Consistent with dust emitting components that are likely co-incident with the
compact gaseous disk detections in NGC\,83 \citep[][]{you05} and NGC\,759 \citep[][]{wik97}, both these
galaxies are  
not resolved in the submm images presented in Figures~\ref{fig:ngc83}
and~\ref{fig:ngc759}.
The DSS2 blue and red optical images of these galaxies are saturated;
and here, the $V-R$ extinction maps of 
\citet[][]{you05}
 are 
compared with SHARC\,II emission maps of the dust. 
The compact size derived from the 
$V-R$ extinction map of
the dust disk in NGC\,83 by \citet[][]{you05}
is smaller in extent than the 350\,$\mu$m
continuum 
beam, consistent with the deconvolved, compact submm size of this galaxy
measured from the presented data, and 
that 
submm emission from the dust-disk of NGC\,83 (or NGC\,759) 
being simply 
unresolvable with SHARC\,II. 

The central
disks in NGC\,083 and NGC\,759 
 could represent a late stage in the merging of two disk
galaxies or accretion of gas-rich material.
The gas could have lost momentum in the merging
process (or since the accretion) and fallen to the center of the
galaxy. If this gas component 
is capable of forming stars, it could produce the high phase density
characteristic of ellipticals.  

\subsection{The Quiescent  or Very-Late-Accretion
Ellipticals UGC\,1503 and NGC\,807}
\label{subsec:1503_807}


Of the ellipticals presented here, UGC\,1503 and NGC\,807 are the only
object that are (1) field
galaxies and (2) have 
no currently identified optical morphological signatures of a merger
or accretion event. Therefore, they are 
referred to as quiescent or very-late-accretion, to mean that if there
has been any possible accretion event during their evolutionary lives,
that happened a very
long time ago, and the galaxies have now settled beyond any such
event.
Being field galaxies also means that there is no near neighbor that 
may have interacted with 
them in the near past.

UGC\,1503 and NGC\,807
have intrinsic 50\% contours of the CO-emission 
and 350\,$\mu$m-continuum fluxes that are 
comparable in extent to and
only slightly smaller than those of the optical.
These galaxies probably have dust that is 
distributed in  
slightly edge-on disks with their
near-side 
in the 
northeast and northwest
for UGC\,1503 and NGC\,807
respectively.
\citet[][]{you02}
 report that in
NGC\,807, 70\% of the CO emission is detected in the south of the
galaxy (see Figures~\ref{fig:ngc807}), suggesting the 
bulk of the disk's content is in the southern parts.


\subsection{The Accretion Virgo-Cluster Dwarf Elliptical NGC\,4476}
\label{subsec:4476}

Despite directed searches, H\,I
has not been detected in the cluster galaxy NGC\,4476 (or NGC\,759)
\citep[e.g.,][]{luc05};
 this is in spite of CO-gas detections in
these galaxies 
\citep[][]{lee91, wik95, wik97, you02}.
It is thought the H\,I may been destroyed by the hot cluster gas or, in
the case or NGC\,4476, by ram-pressure stripping 
\citep[][]{luc05}.

Using $HST$ {\bf V} (F555W) imaging, 
\citet[][]{van95}
showed that NGC\,4476 has a regular, 
``spiral-like'' dust lane
that is about $20''$ in diameter.  
They determined a difference
of 19 degrees between the major axis position angle of the dust lane
and that of the stellar component and used the evidence
in this and a sample or other early galaxies to argue that {\em the dust
lane in 
NGC\,4476} did
not originate from
the stellar component 
but {\em was} instead {\em externally accreted}.  
\citet[][]{tom00}
 confirmed the
existence of the regular dust lane using $HST$ {\bf {\it E(V-I)}} [{\it
    F555W - F814W}] extinction images and showed that NGC\,4476 has
H\,II regions. They also noted a blue unresolved nucleus that
they suggested may come from a young star cluster(s) associated with
the dust system.
\citet[][]{you02}
 detected CO gas in a relaxed disk that is co-spatial with
the dust lane and has a diameter of about $27''$.

Figure~\ref{fig:ngc4476} shows 350\,$\mu$m-continuum contours of dust
emission from NGC\,4476 that 
generally follow the direction of the ellipticity and major axis 
position angle of those of the CO disk and 
SDSS-extinction maps 
except at parts of
the 350\,$\mu$m-continuum 
at 30\% flux levels.
In the eastern part of the
galaxy, the 70\%, 50\%, and 30\% contour levels of the SDSS i-band and
SHARC\,II 350\,$\mu$m (respectively of the stellar and dust
components) intensities co-incide quite well. The SDSS extinction image in this
region all extends to the 30\% SHARC\,II 350\,$\mu$m contour
level. 
However, in the
western part of the galaxy, the correspondence of the contours levels
is not so good, suggesting that the dust and stellar disks
are not in the same plane. Indeed, the SDSS-color map shows less
extinction in the 
western part of the galaxy, as may be the case if the dust disk is towards
the background or inner parts of the galaxy in that region.
The 350\,$\mu$m-continuum 50\% 
contour levels along the major axis of the galaxy are
more extended than the CO and stellar intensities of the same relative
levels. These results suggest that the dust component is separate from the
stellar one, as concluded by \citet[][]{van95} from their $HST$ data of
the central region in this galaxy.
The 
350\,$\mu$m-continuum 30\% contour levels 
(which are not detected at a high signal-to-noise)
are also 
more
 extended than the CO 
and extinction ones and have 
a 
spur 
in the
southwest.
This emission spur may be part of the dust disk in the eastern
background of the galaxy. 


The H\,II regions
 in NGC\,4476 (and NGC\,5666) most probably obtain their fuel from the
 dust seen both in the optical extinction and submm and in turn heat
 the dust to produce the FIR-submm emission seen in these
 galaxies. Futher discussion of star-formation, the heating mechanisms,
 as well as the dust sources are discussed in Section~\ref{sec:sfr_es}
 below and will 
 subsequently be analysed in more detail in a follow-up
paper that uses $Spitzer$ mid-IR spectroscopy of these galaxies (Leeuw
 et al. {\it in prep}).












 
\section{
The Mid-IR
to Submm 
Spectral Energy Distribution 
and Luminosities}\label{sec:sed_es}


The mid-infrared (mid-IR or MIR) flux in ellipticals 
comes primarily from 
``warm'' galactic dust, nebulae, or the envelopes of evolved stars, 
while the FIR to submm flux 
comes 
mainly
from cold dust with temperatures, 
$T \la 100$\,K; 
and, in the case of
radio-bright sources, 
it can also originate  from the high-frequency-radio
components.  On the other hand, flux in the optical/near-IR 
comes
primarily from the stellar components of 
$T \ga 1000$\,K, 
with negligible
contribution to the FIR to submm spectrum.  
For the radio-faint sources observed in this program, 
it is assumed that the mid-IR to submm SED emission comes from 
a cold to cool (5\,K$ < T < 100$\,K) thermal component 
plus a mid-IR power-law component that 
includes PAHs, warm ($T \ga 100$\,K) dust, and stellar galactic components.



Motivated by the general appearance of the SEDs of the elliptical
galaxies and the available data, the SED is modeled with a power law
in the mid-IR (8 to 40\,$\mu$m)
and a graybody in the far-IR (40 to 2000\,$\mu$m).
The Mid-IR power-law fit 
 is not a physical model, but 
rather a mathematical tool to aid in the calculation of total mid-IR luminosity for the galaxies.
However, the graybody fit is an attempt at physically modeling the average temperature, emissivity, and optical depth of the cold dust emission for the galaxies.  

\begin{figure}[h]
\vspace*{-0.6cm}
\begin{center} 
\includegraphics[angle=-90,width=3.215in]{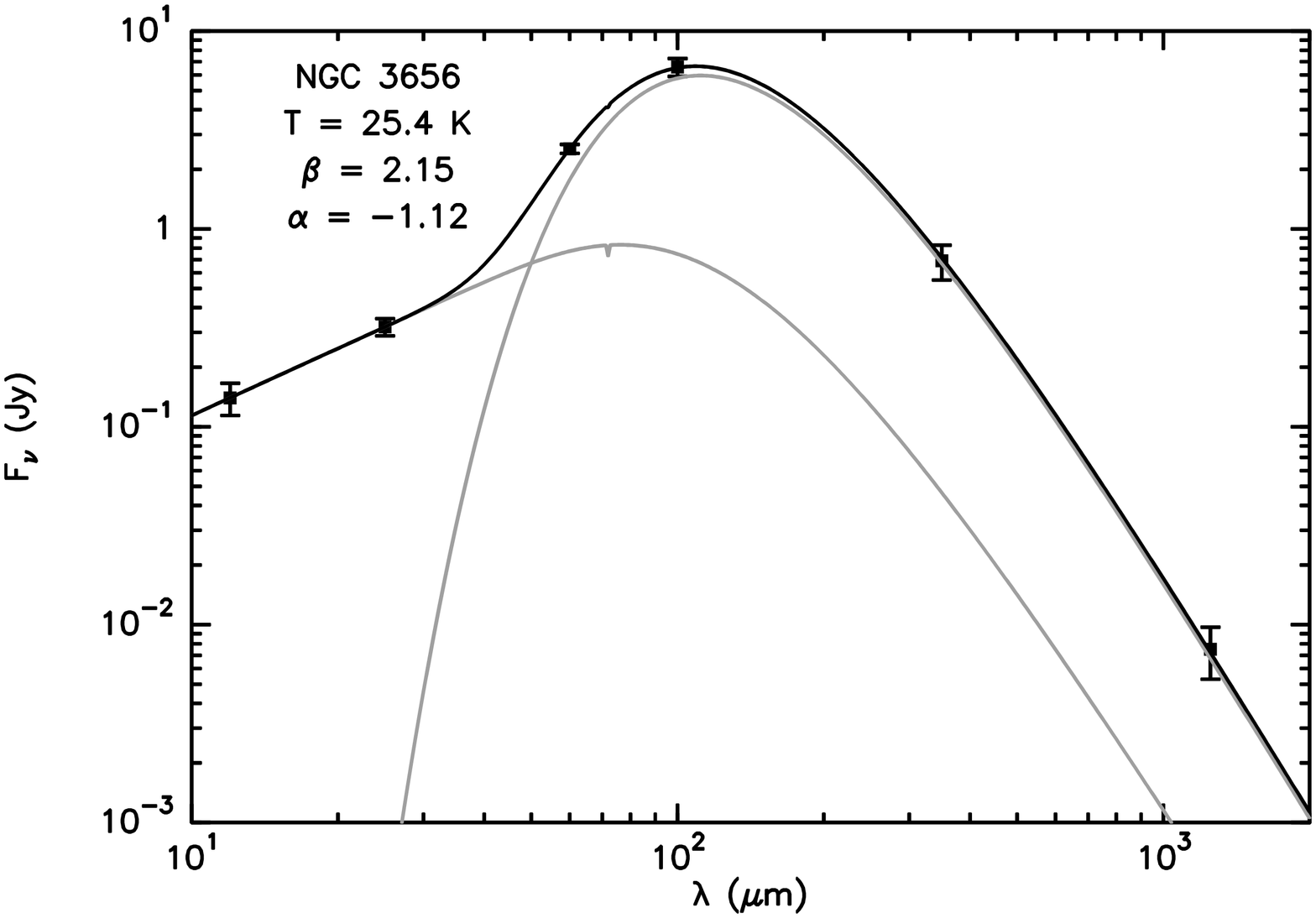}\\
\includegraphics[angle=-90,width=3.215in]{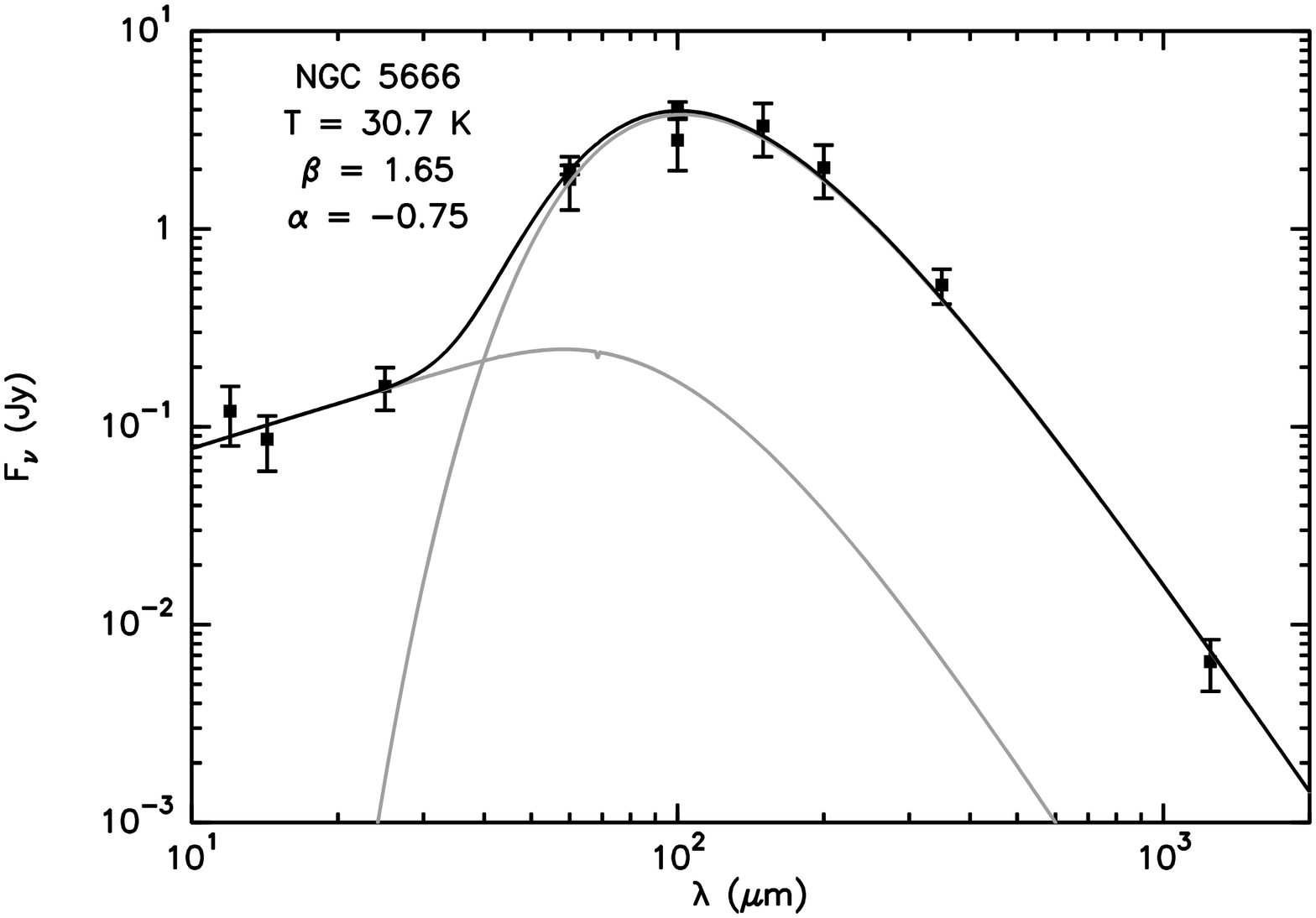}
\includegraphics[angle=-90,width=3.215in]{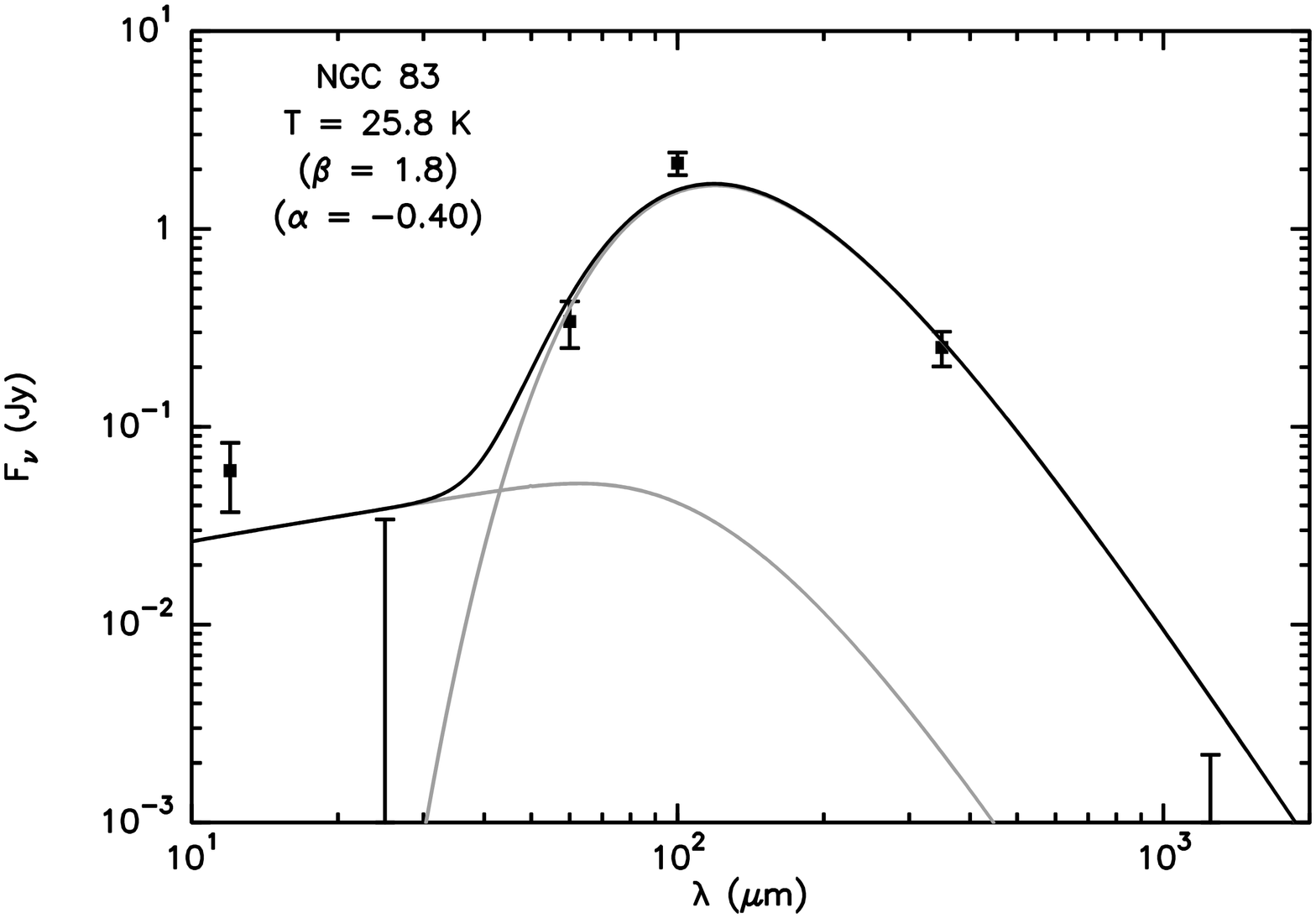}
\includegraphics[angle=-90,width=3.215in]{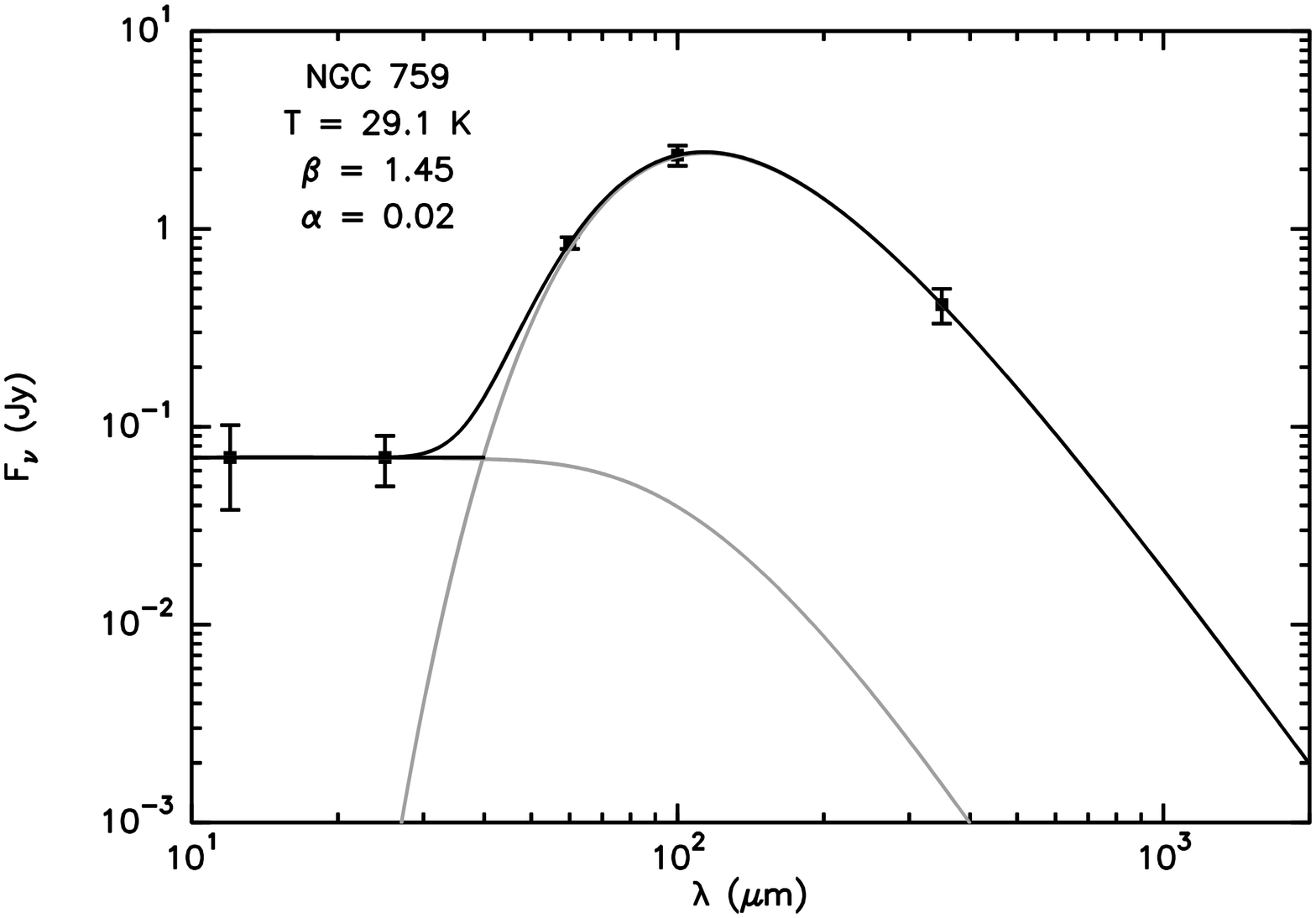}
\includegraphics[angle=-90,width=3.215in]{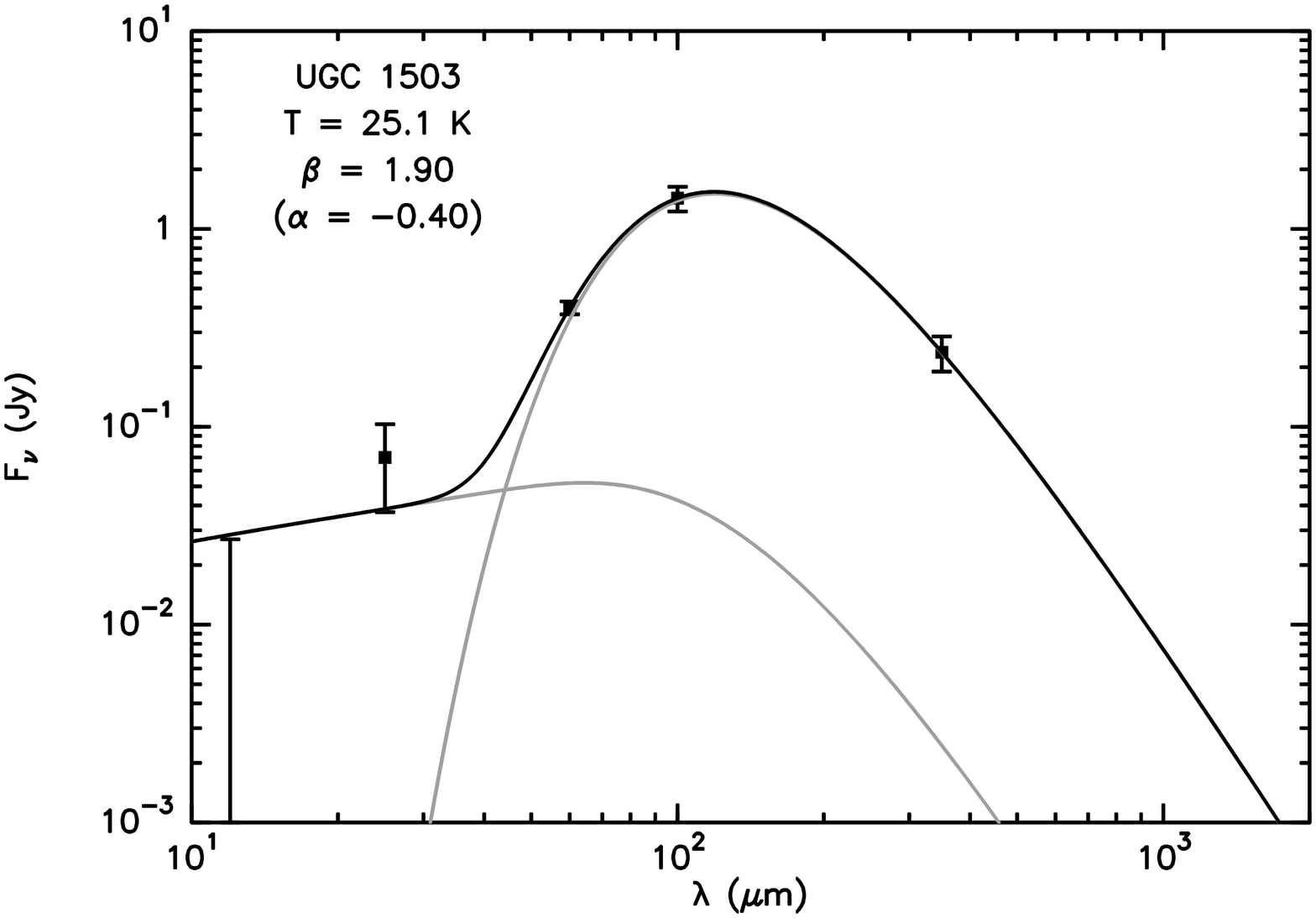}
\includegraphics[angle=-90,width=3.215in]{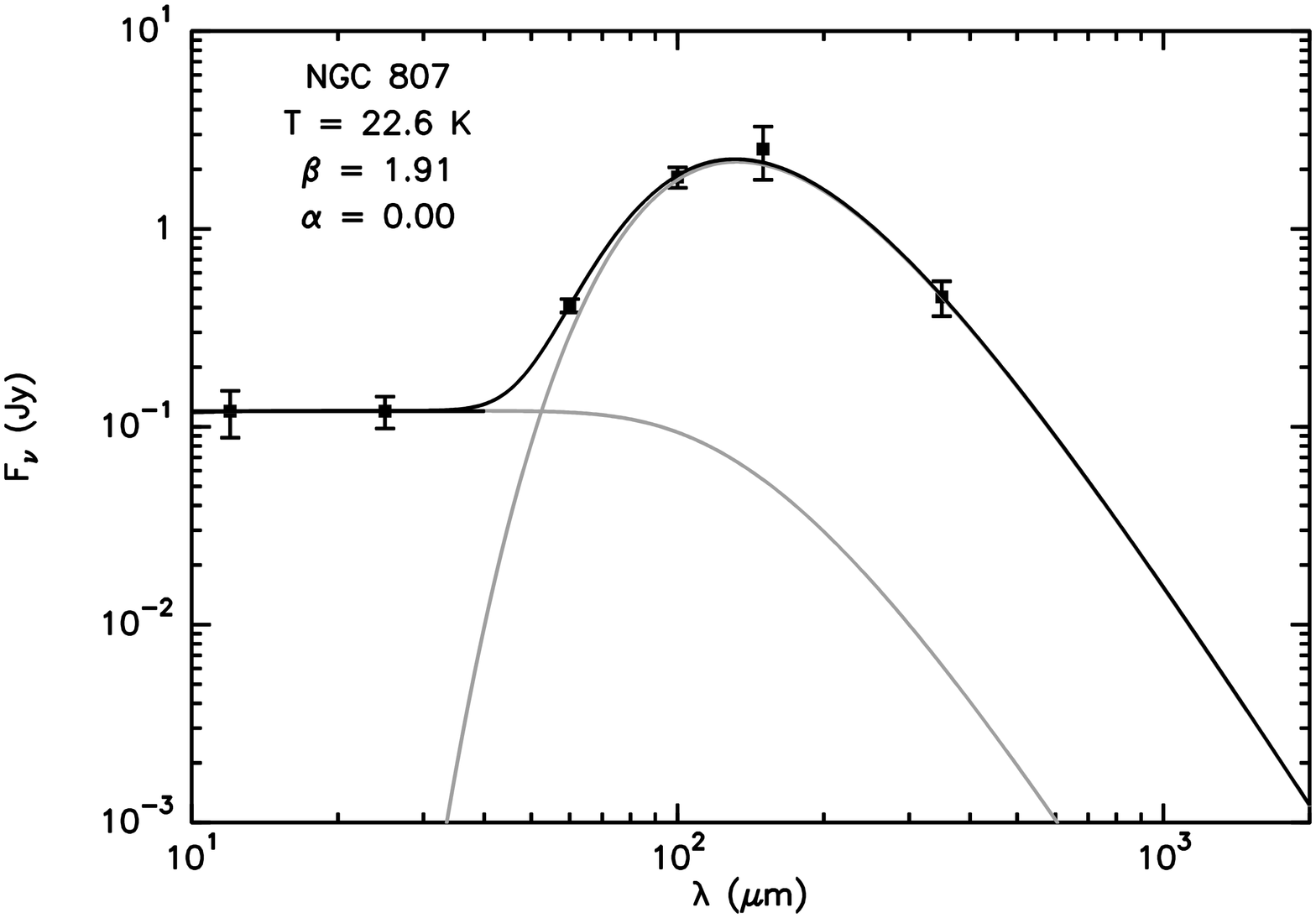}
\includegraphics[angle=-90,width=3.215in]{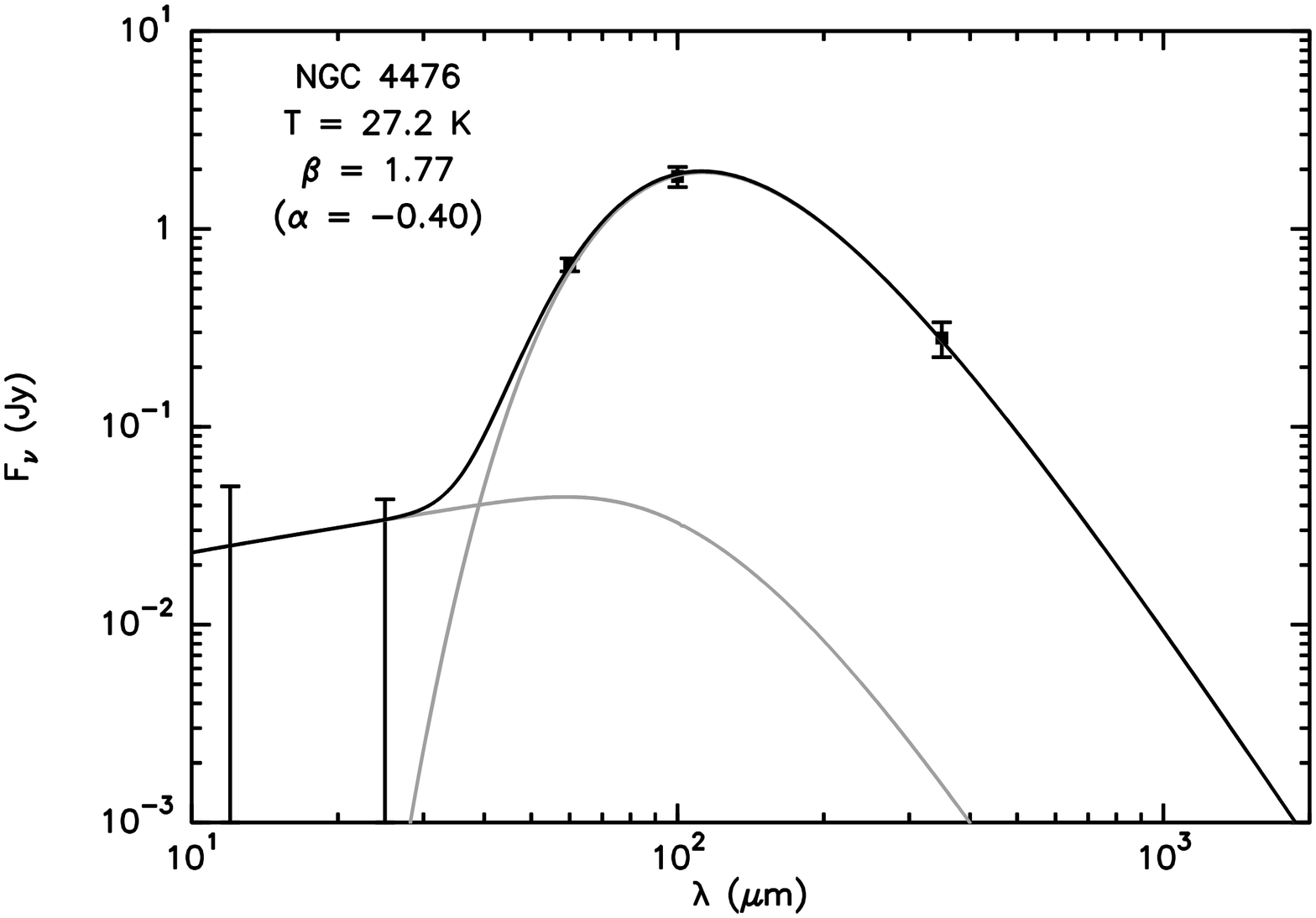}
\vspace{-0.3cm}
\caption{
The CSO/SHARC\,II 350\,$\mu$m continuum
integrated fluxes and mid-IR to
submm data obtained from public archives and literature (see Table~\ref{tab:iras}) for the
sample galaxies are plotted and fitted with a composite model of a
single-temperature greybody plus mid-IR power-law (see Section~\ref{sec:sed_es}). 
\label{fig:cso_sed_bb_pw}}
\end{center}
\vspace*{-0.9cm}
\end{figure}

The model fits are plotted in 
Figure~\ref{fig:cso_sed_bb_pw}, and their
derived parameters are labeled on the plots as well as listed in
Table~\ref{tab:cso_1}.
The power-law model is generated with a truncated $F {\nu}$
temperature distribution, and thus the shape of the mid-IR components
shown in the SED plots.  
For sources which have less than two measurements in the
mid-IR, the median power law 
$F {\nu}^{\sim -0.4} $
is assumed. 
The power law 
for NGC\,4476,
which has no mid-IR detection,
shows a $3\,\sigma$
upper limit in Figure~\ref{fig:cso_sed_bb_pw}.  
Reasonable temperature and beta are derived for the graybody part of
the spectrum for all sources
except NGC\,83.  For that source, it is assumed that the dust emissivity index
beta is equal to the median value of 1.8.









Cold dust 
temperatures and emissivity indices 
are respectively determined to be between 
22\,K $<T<$ 31\,K and 
$1.6 \leq \beta \leq 2.2$ 
for the sample.
These parameters are typical for single-graybody
 fits to
unresolved FIR-to-submm fluxes of dust emitting from elliptical galaxies
\citep[e.g.,][]{lee04, tem04}.
  However, they
are about seven degrees cooler than dust temperatures of ellipticals derived from 60 and
100\,$\mu$m flux ratios alone and a fixed $\beta = 1.0$ (e.g. Wiklind et al. 1995).
On the other hand, the dust temperatures are also slightly warmer and
the emissivity 
indices similar 
to those of dust in the Milky Way Galaxy and nearby spiral galaxies 
\citep[e.g.,][]{dun00}.


The cold dust continuum emission 
presented here extends over
10\,kpc for some galaxies in the
sample,
is detected at high signal to noise with SHARC\,II, and 
is generally co-spatial with previously imaged CO emision
 (see, e.g., Section~\ref{sec:submm_morph}). 
Above the 50\% contour levels, 
the submm dust continuum 
 and optical stellar
intensities of 
similar spatial resolution have the same general
distribution, suggesting that the dust, CO, and stellar components
in the central regions are all co-spatial. The early-merger elliptical
NGC\,3656, with its disturbed optical morphology,
is an exception to this coupling of submm and optical
central flux  
distributions.  

The current maps are not
detected at high enough significance that proper dust distributions can
be discerned below the 30\% contour levels.  If there
exists any dust component for the sample that 
is cooler
or 
has lower-level emission that perhaps
extends over the entire galactic stellar or H\,I distributions  (see, e.g., Section~\ref{sec:sfr_es}),
its detection 
will have to await the next-generation submm detectors such as those
that will be
on ALMA, $Herschel$, 
JCMT, or SOFIA.    
Combining the
current data set together with 
anticipated $Spitzer$ and other upcoming sensitive FIR-to-submm 
observations and analysing the 
SED 
of mapping data from various FIR-to-submm bands 
should put tighter constraints on the physical parameters
of cold dust in the sample. 





MIR-to-submm luminosities ($L_{0.04-1{\rm mm}}$ and $L_{8\mu{\rm m}-1{\rm
    mm}}$) that were obtained by integrating below the model
fits
shown in Figure~\ref{fig:cso_sed_bb_pw}
are listed in Table~\ref{tab:cso_1.5}  
and these can be compared to the optical 
luminosities ($L_{{\rm B}}$) that are also listed in Table~\ref{tab:cso_1.5}.
The percentage excess MIR-to-submm luminosities
over the total luminosities ($L_{{\rm B}-1{\rm mm}}$) 
are
determined and 
also 
listed in Table~\ref{tab:cso_1.5}.  
For most of the galaxies, the MIR-to-submm excess luminosities are
about 20\% to 30\% of the total  luminosities, with about 70\% or more
of 
the excess 
emitted at FIR to mm wavebands
as thermal, cold dust 
emission (see
Figure~\ref{fig:cso_sed_bb_pw} and Table~\ref{tab:cso_1.5}).
The respective early-age and intermediate-age galaxies NGC\,3656 and NGC\,5666 have
higher than 
average MIR-to-submm excesses 
of the total 
luminosities (
$\ga 50$\%)
for the sample (see Table~\ref{tab:cso_1.5}),
suggestive of more or denser dust and star-formation or, 
at least, fairly high 
MIR-to-submm 
emission 
 in comparison to 
total emission
in these more recent merger-remnant/dust-accretion
galaxies.





\section{
Estimates of Star Formation Rates 
and Other Luminosity Sources}
\label{sec:sfr_es}


The peak 
350\,$\mu$m
 optical depths given in 
Table~\ref{tab:cso_3} 
were determined using the 
350\,$\mu$m
 continuum fluxes  and 
 intrinsic 
350\,$\mu$m
 sizes 
respectively presented in Tables~\ref{tab:iras}
and
~\ref{tab:cso_2}
 for each galaxy.
According to
\citet{dra03},
the corresponding
peak optical extinctions are about 14,000 times larger than the sub-mm
optical depth values, implying
peak visual extinctions ($A_V$)
greater than 2\,mag for all the galaxies in the sample.
Therefore, the dust
measured at 350\,$\mu$m for these galaxies
is dense enough to 
absorb almost all the UV or optical radiation
emitted by any massive star embedded within it.

If the far-IR luminosity, L(FIR), for each galaxy in our sample is due to the
reprocessing of radiation from embedded, newly formed stars,
then 
an estimate can be made of the current Star Formation Rate, SFR, in each
galaxy using the relation given in Kennicutt (1998)
\begin{equation}
       SFR ({\rm M}_{\odot} / {\rm yr}) = 1.75 \times 10^{-10} L_{\odot}{\rm(FIR)},
\end{equation}
where L(FIR) includes radiation from 8 through to 1000\,$\mu$m.
Table~\ref{tab:cso_1.5} lists
the estimate of L(FIR) for each galaxy
based on the fits to the data shown in Figure~\ref{fig:cso_sed_bb_pw}.
Except for NGC\,4476,
Equation 1 yields SFR estimates ranging from 1 to 3 ${\rm M}_{\odot}/{\rm yr}$
for the galaxies in the sample (see Table~\ref{tab:cso_3}).
These rates are similar to the SFR in the Milky Way, a
spiral galaxy, but are about 10
times those for `normal' elliptical galaxies.  NGC\,4476 has a SFR
estimate more in line with a normal elliptical galaxy.  However, NGC\,4476 is
exceptional in this sample in that it is a dwarf elliptical
near the center of the Virgo cluster.

For the most part, the FIR luminosity given in Table~\ref{tab:cso_1.5}
for each galaxy is composed of  power radiating at wavelengths
longer than 40\,$\mu$m.
Based on the SED's shown in Figure~\ref{fig:cso_sed_bb_pw}, the
majority of this FIR power, therefore, could be coincident
with the 350\,$\mu$m emission, which seems to be embedded in rotating
CO ``disks''.
Thus, the FIR data and our 350\,$\mu$m data imply for most of the
galaxies in our sample,
the presence of mini-starbursts inside central rotating disks of gas
and dust.

There is another more indirect way to estimate SFR, by
using the empirical relationship from \citet[][]{ken98} between SFR surface density, $SD_{\rm(SFR )}$, and the
molecular gas surface density, $SD_{\rm (Gas)}$: 
\begin{equation}
       SD_{\rm(SFR )}({\rm M}_{\odot}/{\rm yr}/{\rm kpc}^2) = 2.5
       \times 10^{-4} (SD_{\rm (Gas)}({\rm M}_{\odot}/{\rm pc}^2))^{1.4}.
\end{equation}
Estimates of the SFR surface densities at the peaks
of the gas distributions in these galaxies have been derived 
based on the total molecular gas masses (Table~\ref{tab:one}) and the
intrinsic sizes of the CO distributions in these
galaxies (Table~\ref{tab:cso_2}).  The product of the SFR surface densities with the
intrinsic areas of the
CO distributions yield the total SFRs for the galaxies and these are listed in 
Table~\ref{tab:cso_3}.

A comparison between the SFR values estimated using the two methods given above
would imply that NGC\,759 and, especially, NGC\,3656 have the
potential (in gas) for considerably 
more star formation than is presently being observed in the FIR, but that
the other galaxies in our sample are using their gas reservoirs at about the
``normal'' rate to form stars. 

The discussion above assumes the FIR excess from these galaxies
originates from star formation; however, as noted in the Section 1,
\citet[][]{mat03}
 argue that the FIR excess could be due to
the thermal energy released by the hot ISM as it cools through thermal
electron collisions with dust grains. In this scenario, the central
cold dust seen in these elliptical galaxies originates from the
settling of the cooled hot dusty gas from mass losing red giants
distributed throughout each of the elliptical galaxies, 
rather than from the
gas and dust from recent galaxy merger or accretion events.
\citet[][]{mat03}
estimated the FIR luminosity radiated from the dust grains involved in
the cooling of the hot gas of about $10^{7} K$ for a given total
stellar mass-loss rate, $\dot{\rm M}/{\rm M}_{\odot}{\rm yr^{-1}}$, to
be approximately   
\begin{equation}
       L_{\rm(FIR)}({\rm L}_{\odot}) \sim 5.1 \times 10^{7} ({\rm T}/{10^7 K})(\dot{\rm M}/{\rm M}_{\odot}{\rm yr^{-1}}).
\end{equation}
Substituting the FIR luminosity listed for each galaxy in
Table~\ref{tab:cso_1.5} gives mass-loss rates ranging from 15 to 340
${\rm M}_{\odot}{\rm yr^{-1}}$ for this set of galaxies.  These rates
are much higher than those for giant elliptical galaxies, which are
normally $ \sim 1 {\rm M}_{\odot}{\rm yr^{-1}}$.  The larger mass-loss
rates seem consistent with the larger than normal dust masses at the
center of these galaxies as implied by our 350\,$\mu$m data; however,
the optical/near-IR luminosity for each galaxy in the sample presented
here is similar to normal giant elliptical galaxy optical/near-IR
luminosities, so normal stellar mass-loss rates would be expected for
these galaxies. 
Hence, although both the FIR luminosity and the
350\,$\mu$m dust mass estimates are consistent with large stellar
mass-loss rates for the galaxies in this sample, the optical/near-IR
luminosities for these galaxies are not consistent with such a stellar
mass-loss increase. 

In summary, the calculations given in this section favor the FIR
luminosity of these galaxies resulting primarily from 
cold dust in the central currently star formation regions 
that were probably induced and supported by accretion or merger events
involving a gas- or dust-rich galaxy or satelite, 
rather than 
coming from hot ISM cooling in collisions 
with dusty gas from stellar mass-loss.
However, a small FIR luminosity contribution from the hot ISM cooling,
the stellar, or any other dilute galactic radiation field is not excluded.



\section{Dust Masses 
and Gas-to-Dust Mass Ratios} 
\label{sec:dustmasses} 

Following 
\citet[][]{hil83}, the mass of dust, $M_{\rm d}$, that emits mid-IR to submm continuum can 
be estimated from
\begin{equation}
M_{\rm d}= \frac{{F_{\nu} D^2}}{{k_{\rm d} B(\nu ,T)}},
\label{eqn:md}
\end{equation}
\noindent
where $F_{\nu}$ is the measured flux density at frequency $\nu$, $D$ is the
distance to the source, $B({\nu} ,T)$ the Planck function and $k_{\rm d} =
3Q_{\nu}/4a\rho$ the grain mass absorption coefficient where $a$ and
$\rho$ are respectively the grain radius and density.  A recently
updated value of
$k_{\rm d}^{350\,\mu {\rm m}} = 1.92 \times 10^{-1}\,{\rm m}^2 {\rm kg}^{-1}$
\citep[][]{dra03}, as for Galactic
 dust (cf. above),
is assumed, yielding dust masses that range
from $\sim 9
\times 10^5$ 
to $
\sim 2.5 \times 10^7 {\rm M}_{\odot}$ for 
 dust with temperatures 
in the range 22\,K$ \la T \la 30$\,K
(see Figure~\ref{fig:cso_sed_bb_pw} and 
Table~\ref{tab:cso_1}).  




NGC\,4476 has the lowest dust mass in the sample. The relative paucity
of dust in this galaxy is not surprising as NGC\,4476 is a dwarf
elliptical.  However, 
\citet[][]{luc05} 
 observed that this
galaxy, which lies within the Virgo Cluster, has been stripped of its
H\,I, maybe through ram-pressure; therefore, the presence of any compact
distribution of CO and cold dust at all implies that the H\,I is more
vulnerable to ram-pressure stripping than is the denser molecular gas
and the dust associated with it
within this cluster
galaxy. 

%


Table~\ref{tab:cso_1} shows that the sample has diverse H$_2$-mass to dust-mass ratios 
that are mostly between
$120 \la M$(H$_2$)/$M_{\rm dust} \la 160$,
and in the quiescent galaxy NGC\,807 and early-merger remnant NGC\,3656 respectively range
from $\sim 60$ (about half the average) to $\sim
310$  (about twice the average). 
The ratios suggest that, in the central galactic regions where the H$_2$
and dust masses are respectively measured from the resolved CO and our
CSO continuum measurements,
some of the sample galaxies are 
relatively under-massive in H$_2$
and over-massive in dust content, while the 
others are over-massive in H$_2$ and under-massive in dust content.
The implications of these mass ratios
and their possible 
association with star-formation 
or, at least, gas-phases in the central regions of
these galaxies 
will be investigated further in our upcoming paper that
uses $Spitzer$ IRS data to 
study the MIR spectral evolution and/or 
star-formation properties of
these galaxies.

The total cold-gas-mass (i.e., (H$_2$ + H\,I)-mass) to
dust-mass ratios for the sample are, however, less diverse 
and in 
a narrower range of 
$230 \la M$(H$_2$ + H\,I)/$M_{\rm dust} \la 400$, with 
NGC\,807 and NGC\,3656 
at much closer ratios of 300 and 387,
respectively (see Table~\ref{tab:cso_1}).
At face value, it may appear that the 
total cold-gas-mass
to dust-mass ratio for the sample
is
about 3 times the value for the Milky Way Galaxy (i.e., $M$(H$_2$ +
H\,I)/$M_{\rm dust} \sim 100$). 
However, 
in sample objects where H\,I has been
resolved \citep[e.g.,][]{lak87, bal01}, the atomic gas  has been observed to be much more
extended than the inner galactic regions 
where our observations have mapped dust emission and shown it to be
associated with molecular gas.
A cold and extended dust component may yet be detected that 
is associated with the atomic hydrogen, and perhaps massive enough to
bring the total cold gas-to-dust mass ratio for the sample in line
with the Milky Way value.

\section{Summary of Results and 
Conclusions
}\label{sec:sum_es}  

SHARC\,II continuum observations at 350\,$\mu$m
have been presented of 
seven elliptical galaxies 
with known CO-gas disks 
\citep[][]{wik97, you02, you05}.
The SHARC\,II mapping 
provides the first {\em clearly resolved}
FIR-submm 
continuum emission from cold dust (with temperatures $\sim 31$\,K $> T >$
23\,K) of any elliptical galaxy $>40$\,Mpc.       
Calculations of the measured FIR excess show that the dust's
most likely and dominant heating source 
is not dilute stellar radiation or cooling flows but rather
star-formation, that 
could have been 
triggered by an accretion or
merger event and fueled by dust-rich
material 
that has settled in a 
dense region 
co-spatial with the
central CO-gas disks.
The 
dust is detected even in two 
cluster ellipticals
that are deficient in H\,I, showing 
that, unlike the H\,I, cold dust and CO in ellipticals can survive among
hot X-ray gas, even in galaxy clusters.

Above the 50\% contour levels, where 
the submm detections are of high significance, the submm dust continuum 
 and optical stellar
intensities of 
similar spatial resolution have the same general
distribution, suggesting that the dust, CO, and stellar components
in the central regions are co-spatial. The early-merger elliptical
NGC\,3656, with its dense optical dust lane that is thought to be in a highly
inclined warped disk, is an exception to this coupling of submm and optical
central flux  
distributions.  Below the 50\% contour levels,
and especially below the 30\% levels, the current maps are not
detected at high enough significance that proper dust distributions can
be discerned.  

No dust cooler than 20\,K, distributed outside the CO disks, 
or co-spatial with and heated by the entire, dilute stellar optical galaxy
(or very extended H\,I) 
is currently evident.
These observations clearly show that the
FIR-submm emission in our sample ellipticals 
is primarily from cold dust
that is associated with star-formation 
located in the central parts of these
galaxies.      
The results support the hypothesis that if ellipticals are formed
through mergers of gas-rich spirals, their 
starbursts (1) are necessary to produce the high phase-space
density seen in the center of ellipticals 
and (2) would power the FIR-bright stage as the merger relaxes. 
Combining the
current data set together with 
anticipated $Spitzer$ and other upcoming sensitive FIR-to-submm 
mapping observations 
may detect or
rule-out the existence of larger-scale, diffusely distributed cold 
dust with an average
temperature that is perhaps $<20$\,K (and maybe associated with H\,I).

Assuming the grain mass absorption coefficient 
$k_{\rm d}^{350\,\mu {\rm m}} = 1.92 \times 10^{-1}\,{\rm m}^2 {\rm kg}^{-1}$,
as for Galactic-like dust 
\citep[][]{dra03}, 
the
measured submm fluxes and dust temperatures yield dust masses for the
sample that range
from $\sim 9
\times 10^5 {\rm M}_{\odot}$ to $
\sim 2 \times 10^7 {\rm M}_{\odot}$.  
The sample has diverse H$_2$-mass to dust-mass ratios 
that 
cover
$60 \la M$(H$_2$)/$M_{\rm dust} \la 310$,
suggesting that in the central galactic regions, where the H$_2$
and dust masses are respectively measured from the resolved CO and our
CSO/SHARC\,II continuum measurements,
some of the sample galaxies are 
relatively under-massive in H$_2$
and over-massive in dust content, while others are the opposite.
The possible implications of these mass ratios
on star-formation 
or, at least, gas-phases in the central regions of
these galaxies 
will be investigated further in our upcoming paper that
uses $Spitzer$ IRS data to 
study the detailed MIR spectral 
and/or 
star-formation 
properties of
these galaxies. 

The total cold-gas-mass 
to dust-mass ratios for the sample are
in 
a narrower range of 
$230 \la M$(H$_2$ + H\,I)/$M_{\rm dust} \la 400$, 
with a total cold-gas-mass
to dust-mass ratio for the sample
that, at face value, may appear to be
about 3 times the value for the Milky Way Galaxy (i.e., $
\sim 100$). 
However, 
in sample objects where 
atomic gas 
has been
resolved \citep[e.g.,][]{lak87, bal01}, 
the H\,I
has been observed to be much more
extended than the inner galactic regions 
where our observations have mapped dust emission 
that is associated with molecular gas.
A cold and extended dust component may yet be detected that 
is associated with the atomic hydrogen, and perhaps massive enough to
bring the total cold gas-to-dust mass ratio for the sample in line
with the Milky Way value.
As noted above, this 
may be oberved with future more sensitive FIR/submm 
detectors
such as those that will be on SOFIA, $Herschel$, 
JCMT, or ALMA. 

\acknowledgments

Part of this work was done while Lerothodi L. Leeuw (LLL) was at the
University of Chicago and partially supported by a Chicago 2004 NASA
Mini-Award Grant and NSF grant AST-0505124 and at NASA Ames Research
Center.  LLL finalized this research and wrote this paper while  
supported by the South Africa
Square Kilometer Array Postdoctoral Bursary at Rhodes University.
Research at the Caltech Submillimeter Observatory is supported by NSF
grant AST-0540882.

This publication makes use of data products from the second Digital Sky
Survey (DSS2), that was based on photographic data obtained using
the Oschin Schmidt Telescope on Palomar Mountain. The Palomar
Observatory Sky Survey was funded by the National Geographic
Society. The Oschin Schmidt Telescope is operated by the California
Institute of Technology and Palomar Observatory. The plates were
processed into the present compressed digital format with their
permission.  The DSS was produced
at the Space Telescope Science Institute under US Government grant NAG W-2166.

\clearpage

\begin{table}[ht!]
\centering
\caption[]{
\small{ 
Published optical and molecular-gas properties of the 
ellipticals 
\citep[adopted from][and references therein]{mur00, tom00, wik95b,
  wik97, you02, you05}
}}
\label{tab:one}
\vspace*{0.1cm}
\begin{tabular}{lccccccc}
\hline
Source & Type & Environment, Notables, \& &
$M\rm{(H_2)}$   & 
$M\rm{(HI)}$   & 
$D$ & $L_B$ & $(B-$ \\
& & Merger or Accretion Stage & 
[$10^8$ & 
[$10^8$ & 
[M & 
[$10^9$ & $V)_e$
 \\ 
& & &
${\rm M}_{\odot}$] & 
${\rm M}_{\odot}$] &  
pc] & 
${\rm L}_{\odot}$] &  mag
 \\ 
\hline
NGC 3656  & Ep & Early-Age Major Merger/Accr.? & 47 
& 11 & 45 & 16 & \\
NGC 5666  & cE2/SO & Intermediate Merger/Accr.? & 5.7 
& 10 & 35 & 6 & 0.86 \\
NGC  83  & E0 & Group Member, Late Merger/Accr.? 
&  20
& & 85 & 44 & 1.12  \\
NGC  759  & E0/1 & Cluster A262, 
Late Merger/Accr.? & 24 
& $<$24 & 66 & 34 & 1.05 \\
NGC  807  & E3 & Field, Quiescent / Very-Late Accr.?
& 14 
& 61 & 64 & 31 & 0.97 \\
UGC 1503  & E1 & Field, Quiescent / Very-Late Accr.?   & 18 
& 16 & 69 & 16 & \\
NGC 4476  & dE5p/SO
& 
Virgo Cluster, Dwarf Merger/Accr.?
& 1.1 
& $<$1 & 18 & 3 & 0.85  \\
\hline
\end{tabular}
\end{table}

\begin{table}[h!]
\vspace{-0.3cm}
\centering
\caption[]{
\small{
SHARC\,II \& $IRAS$ 
continuum fluxes for the selected elliptical galaxies 
}}\label{tab:iras}
\begin{tabular}{lcccccc}
\hline
Source  & Peak @350$\mu$m & Total & Total & Total & Total & Total \\
  &  (mJy/$9''$ beam) & 350$\mu$m (Jy) & 100$\mu$m (Jy) & 60$\mu$m (Jy) & 25$\mu$m (Jy) & 12$\mu$m(Jy) \\
\hline
NGC 3656   & $ 368 \pm 73$ &  $0.69 \pm 0.14$ & $6.58 \pm 0.68$ & $2.54 \pm 0.13$ & $0.32 \pm 0.03$ & $0.14 \pm 0.03$  \\
NGC 5666   & $ 150 \pm 30 $ & $0.52 \pm 0.10$ & $3.98 \pm 0.40$ & $1.99 \pm 0.10$ & $0.16 \pm 0.04$ & $0.12 \pm 0.04$  \\
NGC  83   & $ 191 \pm 38 $ & $0.25 \pm 0.05$ & $2.15 \pm 0.28$ & $0.34 \pm 0.09$ & $ <0.03 $ & $0.06 \pm 0.02$  \\
NGC  759   & $ 71 \pm 14 $ & $0.42 \pm 0.08$ & $2.36 \pm 0.28$ & $0.85 \pm 0.06$ & $0.07 \pm 0.02$ & $0.07 \pm 0.03$  \\
NGC  807   & $ 63 \pm 12 $ & $0.45 \pm 0.09$ & $1.83 \pm 0.22$ & $0.41 \pm 0.03$ & $0.12 \pm 0.02$ & $ 0.12\pm 0.03$  \\
UGC 1503   & $ 57 \pm 11 $ & $0.24 \pm 0.05$ & $1.43 \pm 0.20$ & $0.40 \pm 0.03$ & $0.07 \pm 0.03$ & $<0.03$ \\
NGC 4476   & $ 117 \pm 23 $ & $0.28 \pm 0.06$ & $1.84 \pm 0.21$ & $0.66 \pm 0.05$ & $<0.04$ & $<0.05$  \\
\hline
\end{tabular}
\end{table}

\begin{table}[ht!]
\vspace*{-0.3cm}
\centering
\caption[]{
\small{ 
Intrinsic 
Galactic sizes of the stellar, molecular gas, and dust
distributions at the 50\% contour levels of 
the 
respective peak
surface brightness 
(The listed SDSS i-band size for NGC\,3656 is at the 70\% contour level. See text for further details.)
}}
\label{tab:cso_2}
\vspace*{0.1cm}
\begin{tabular}{lcccccc}
\hline
\multicolumn{1}{c}{Source} & \multicolumn{2}{c}{@50\% 350\,$\mu{\rm
m}$ Flux} & \multicolumn{2}{c}{@50\% CO Flux} &
\multicolumn{2}{c}{@50\% SDSS i-Band Flux} \\
\multicolumn{1}{c}{} & \multicolumn{2}{c}{} & \multicolumn{2}{c}{} &
\multicolumn{2}{c}{{\it or} $^*$ @50\% DSS2 Red Flux} \\
 & arcs $\times$ arcs & kpc $\times$ kpc
& arcs $\times$ arcs & kpc $\times$ kpc
& arcs $\times$ arcs & kpc $\times$ kpc \\
\hline
NGC 3656  & 
$6.5 \times 12.7$ & $ 1.5 \times 2.8$
& $2.6 \times 13.2$ & $ 0.5 \times 2.8$
 & 
$11.2 \times 17.3$ &  $2.4 \times $ 3.8 \\
NGC  5666  & 
$10.3 \times 11.2$ & $ 1.8 \times 2.0 $
& $8.2 \times 11.1$ & $ 1.3 \times 1.8 $
 & 
$12.5 \times 15$ & $ 2.1 \times 2.5 $ \\
NGC  83  & 
$4.1 \times 5.0$ & $ 2.0 \times 1.8 $
& $3.8 \times 5.2$ & $ 1.5 \times 2.0 $
& 
{\it saturated} & {\it saturated} 
\\
NGC  759  & 
$5.0 \times 5.0$ & $ 1.5 \times 1.5 $
& $3.7 \times 4.5$ & $ 1.0 \times 1.5 $
& 
{\it saturated} & {\it saturated} \\
NGC  807  & 
$13.7 \times 30.4$ & $ 4.3 \times 9.3 $
& $11.7 \times 20.3$ & $ 3.5 \times 6.3 $
& 
$^*20 \times 32$ & $^*6.3 \times 9.8 $ \\
UGC 1503  & 
$12.1 \times 21.4$ & $ 4.0 \times 7.0 $
& $9.4 \times 15.2$ & $ 3.0 \times 5.0 $
& 
$^*15.3 \times 18.9$ & $^*5.0 \times 6.3 $ \\
NGC  4476  & 
$4.1 \times 19.1$ & $ 0.3 \times 1.5 $
& $3.0 \times 13.4$ & $ 0.3 \times 1.0 $
 & 
$2.0 \times 9.8$ & $0.2 \times 0.9$ \\
\hline
\end{tabular}
\end{table}

\begin{table}[ht!]
\vspace*{-0.3cm}
\centering
\caption[]{
\small{ 
Mid-IR to SHARC\,II-derived 
cold-dust parameters
and related gas properties 
}}
\label{tab:cso_1}
\vspace*{0.1cm}
\begin{tabular}{lccccc}
\hline
Source 
& $T$ & $\beta$ 
&${M{\rm{(d)}}}$ 
& $\frac{M\rm{(H_2)}}{M{\rm{(d)}}}$ 
& $\frac{M\rm{(H_2 + H\,I)}}{M{\rm{(d)}}}$ \\
& [K] & & [$10^6\,{\rm M}_{\odot}$] & & \\
\hline
NGC\,3656  
& 25.4 & 2.2 & 15 & 313 & 
387 \\
NGC\,5666  
& 30.7 & 1.6 & 4.8 & 119 & 
327 \\
NGC\,83  
& 25.8 & (1.8) & 19 & 105 & \\
NGC\,759  
& 29.1 & 1.5 & 15 & 160 & 
$<$320 \\
NGC\,807  
& 22.6 & 1.9 & 25 & 56 & 
300 \\
UGC\,1503  
& 25.1 & 1.9 & 12 & 150 & 
283 \\
NGC\,4476  
& 27.2 & 1.8 & 0.9 & 122 & 
$<$233 \\
\hline
\end{tabular}
\end{table}

\begin{table}[ht!]
\vspace*{-0.3cm}
\centering
\caption[]{
\small{ 
Mid-IR to mm
luminosity excess
}}
\label{tab:cso_1.5}
\vspace*{0.1cm}
\begin{tabular}{lccccc}
\hline
Source 
& $L^{0.04{\rm mm}}_{{\rm to} \,\,1{\rm mm}}$ & $L^{8\mu{\rm m}}_{{\rm
to} \,\,1{\rm mm}}$ 
& $L^{B \,\,{\rm to}}_{1{\rm mm}}$ & \%$L^{B \,\,{\rm to}}_{1{\rm mm}}$ 
& \%$L^{B \,\,{\rm to}}_{1{\rm mm}}$ \\
& [$10^9$ & [$10^9$ 
&  [$10^9$ & 
as & as \\
& ${\rm L}_{\odot}$] & ${\rm L}_{\odot}$] &
${\rm L}_{\odot}$] 
& $L^{8\mu{\rm m}}_{{\rm to} \,\,1{\rm mm}}$ & $L^{0.04{\rm mm}}_{{\rm to} \,\,1{\rm mm}}$ \\
\hline
NGC\,3656  
& 14.0 &  17.8 
& 33.8 & 53 & 41 \\
NGC\,5666  
& 5.4 & 6.8 
& 12.8 & 53 & 42 \\
NGC\,83  
& 11.7 & 13.8 
& 57.8 & 24 & 20 \\
NGC\,759  
& 10.9 & 13.8 
& 47.8 & 29 & 23 \\
NGC\,807  
& 8.1 & 12.6 
& 43.6 & 29 & 18 \\
UGC\,1503  
& 6.9 & 8.3 
& 24.3 & 34 & 28 \\
NGC\,4476  
& 0.6 & 0.7 
& 3.7 & 19 & 16\\
\hline
\end{tabular}
\end{table}

\begin{table}[ht!]
\vspace*{-0.5cm}
\centering
\caption[]{
\small{ 
The optical depth and total FIR and CO star-formation rate estimates
}}
\label{tab:cso_3}
\vspace*{0.1cm}
\begin{tabular}{lcccc}
\hline
Source &
350\,$\mu{\rm m}$ & A$_{\rm V}$
&  Total & Total \\
& Peak & 
&  FIR SFR & CO SFR \\
& $\tau_{\rm optical}$ & & ${\rm M}_{\odot}$/yr &  ${\rm M}_{\odot}$/yr \\
\hline
NGC\,3656  & 
0.00119 & 17 
& 3.04  & 26.89  \\
NGC\,5666  & 
0.00050 & 7 
& 1.17  & 1.17  \\
NGC\,83  & 
0.00171 & 24
& 2.65  & 6.13 \\
NGC\,759  & 
0.00206 & 29 
& 2.41 & 10.38  \\
NGC\,807  & 
0.00018 & 25
& 2.17 & 1.73  \\
UGC\,1503  & 
0.00015 & 2 
& 1.44 & 2.83  \\
NGC\,4476  & 
0.00047 & 7
& 0.14  & 0.28  \\
\hline
\end{tabular}
\end{table}

\clearpage


\clearpage
\thispagestyle{empty}
\setlength{\voffset}{-18mm}

\clearpage
\setlength{\voffset}{0mm}

\end{document}